\documentclass[a4paper,12pt]{article}
\usepackage[DIV=14,BCOR=0mm,headinclude=true,footinclude=false]{typearea}
\usepackage[T1]{fontenc}

\usepackage{amsmath,amssymb,float,mathrsfs,mathtools,style}

\usepackage{subcaption}
\usepackage{tikz}
\usetikzlibrary{calc,snakes}

\makeatletter
\let\originalleft\left
\let\originalright\right
\renewcommand{\left}{\mathopen{}\mathclose\bgroup\originalleft}
\renewcommand{\right}{\aftergroup\egroup\originalright}
\makeatother

\newcommand{\ed}{\mathop{}\!\mathrm{d}}

\def\be{\begin{equation}}
	\def\ee{\end{equation}}
\def\ba{\begin{aligned}}
	\def\ea{\end{aligned}}

\numberwithin{equation}{section}
\numberwithin{table}{section}

\title{Generalized Couch--Torrence inversions}

\author[a,b]{Panagiotis Charalambous\footnote{\texttt{pcharala@sissa.it}}}
\author[a,b]{Laura Donnay\footnote{\texttt{ldonnay@sissa.it}}}
\author[c,d]{Alexandru Lupsasca\footnote{\texttt{alexandru.v.lupsasca@vanderbilt.edu}}}

\affiliation[a]{International School for Advanced Studies (SISSA), \\
Via Bonomea 265, 34136 Trieste, Italy}
\affiliation[b]{National Institute for Nuclear Physics (INFN), \\
Sezione di Trieste, Via Valerio 2, 34127, Italy}
\affiliation[c]{OpenAI}
\affiliation[d]{Department of Physics \& Astronomy, \\
Vanderbilt University, Nashville TN 37212, USA}
\abstract{
The Laplace equation on Euclidean flat space admits a discrete radial inversion symmetry.
In 1983, Couch and Torrence (CT) found---surprisingly---that the massless wave equation continues to display this symmetry on the background of an extremal (and asymptotically flat) black hole, where the inversion interchanges horizon and infinity while preserving the singularity structure of the separated radial mode equation.
We revisit this CT inversion symmetry and investigate its possible extensions beyond the extremal (Reissner--Nordström or Kerr) setting in which it was originally identified.
Using the example of the static lukewarm de Sitter black hole, we show that neither the exchange of horizon with infinity, nor the preservation of radial singularities, are essential features needed for a CT inversion to exist.
Instead, we interpret CT transformations through their action on photon spheres, providing a unified viewpoint that extends to the (phase-space-dependent) CT inversions of the extremal Kerr–Newman geometry.
For scalar fields on that spacetime, we find a simple relation between the fixed point of a CT inversion and the coefficient of superradiant scattering.
Finally, we exhibit a hidden CT inversion symmetry that arises in the static limit of the Kerr Laplacian for all spins.
Together, these results suggest that CT symmetry may admit a broader generalization than previously understood.
}

\begin{document}

\maketitle

\section{Introduction}
\label{sec:Intro}

The Laplace equation $\nabla^2\Phi=0$ on Euclidean flat space has long been known to admit a discrete radial inversion symmetry ($r\to1/r$) that maps solutions to solutions.
Rather surprisingly, Couch \& Torrence (CT) found in 1983 \cite{Couch1984} that for an extremal (asymptotically flat) black hole, there still exists an exact radial inversion symmetry of the wave equation.
As we now review, for an extremal Reissner--Nordström (ERN) black hole, this ``CT inversion'' is not only a discrete symmetry of the wave equation, but also a conformal isometry of the spacetime itself.

In Schwarzschild coordinates $(t,r,\theta,\phi)$, an ERN black hole of ADM mass $M$ has a degenerate event horizon at $r_+=M$.\footnote{Throughout, we use geometrized units with $G_{\rm N}=c=1$.}
Letting $\ed\Omega_2^2\equiv\ed\theta^2+\sin^2{\theta}\ed\phi^2$ denote the round metric on the unit-radius $2$-sphere, the ERN geometry has line element
\be
	ds^2 = -\frac{\left(r-M\right)^2}{r^2}\ed t^2+\frac{r^2\ed r^2}{\left(r-M\right)^2}+r^2\ed\Omega_2^2 \,.
\ee
Besides time-translation invariance and spherical symmetry, the ERN geometry does not admit any other continuous (exact or conformal) isometries, but it does however possess a \textit{discrete} conformal isometry, which is given by the radial CT inversion~\cite{Couch1984}
\be\label{eq:ERNCT}
	r-M \xrightarrow{\text{CT}} \tilde{r}-M = \frac{M^2}{r-M}
	\quad\Longrightarrow\quad
	ds^2 \xrightarrow{\text{CT}} d\tilde{s}^2 = \Omega^2ds^2 \,,
\ee
with conformal factor $\Omega=\frac{\tilde{r}}{r}=\frac{M}{r-M}=\frac{\tilde{r}-M}{M}$.\footnote{In fact, this CT inversion is an exact isometry of the conformally compactified exterior geometry, $r^{-2}ds^2 = \tilde{r}^{-2}d\tilde{s}^2$~\cite{Borthwick:2023ovc}, and it realizes a self-duality between the horizon and null infinity of the ERN geometry~\cite{Agrawal:2025fsv}.
In tortoise coordinates, the CT inversion takes the very simple form $r_\ast\to\tilde{r}_\ast=-r_\ast$.}
This conformal isometry must induce a symmetry for any massless scalar field $\Phi$ that is conformally coupled to gravity.
Since the ERN metric has vanishing Ricci scalar curvature, the equation of motion for such a probe field is simply the massless Klein--Gordon equation $\nabla^2\Phi=0$.
Under the usual decomposition into monochromatic spherical harmonic modes labeled by frequency $\omega$, azimuthal number $m$ and orbital number $\ell$, $\Phi_{\omega\ell m}(t,r,\theta,\phi)=e^{-i\omega t}R_{\omega\ell m}(r)Y_{\ell m}(\theta,\phi)$, this wave equation reduces to the radial ODE
\be
	\left[\frac{d}{dr}\left(r-M\right)^2\frac{d}{dr}+\frac{\omega^2r^4}{\left(r-M\right)^2}-\ell\left(\ell+1\right)\right]R_{\omega\ell m}(r)=0\,.
\ee
One can check that this ODE is indeed invariant under the conformal transformation
\be
	r-M \xrightarrow{\text{CT}} \tilde{r}-M = \frac{M^2}{r-M} \,,\quad
	R_{\omega\ell m} \xrightarrow{\text{CT}} \tilde{R}_{\omega\ell m} = \frac{M}{r-M}R_{\omega\ell m} \,.
\ee
Thus, the ERN black hole displays a radial inversion conformal symmetry that maps solutions with definite frequency and azimuthal angular momentum $(\omega,m)$ into new solutions with the same mode numbers.
In other words, the wave equation on ERN exhibits an exact conformal symmetry that acts within each superselection sector of fixed energy and angular momentum $(\omega,m)$.

For an extremal Kerr--Newman (EKN) black hole with mass $M$, electric charge $Q$, and angular momentum $J=aM$ saturating the extremal bound $a^2+Q^2=M^2$, this discrete conformal symmetry is no longer present in the spacetime geometry itself.
Nevertheless, a remnant of CT symmetry happens to persist at the level of the (conformally coupled) wave equation $\nabla^2\Phi=0$.
Under the analogous decomposition $\Phi_{\omega\ell m}(t,r,\theta,\phi)=e^{-i\omega t+im\phi}R_{\omega\ell m}(r)S_{\omega\ell m}(\theta)$ into monochromatic \textit{spheroidal} harmonic modes, the wave equation separates into an angular ODE for modes $S_{\omega\ell m}$ (whose pole-regularity fixes the separation constant $A_{\omega\ell m}$), together with a radial ODE
\be\ba
	\left[\frac{d}{dr}\left(r-M\right)^2\frac{d}{dr}+\frac{\left(r^2+a^2\right)^2}{\left(r-M\right)^2}\left(\omega-\frac{ma}{r^2+a^2}\right)^2 -A_{\omega\ell m} \right]R_{\omega\ell m} = 0 \,.
\ea\ee

This equation is left (conformally) invariant under the CT inversions~\cite{Couch1984}\footnote{These CT inversions are conformal symmetries not only of the ``on-shell'' radial ODE (with $A_{\ell m\omega}$ fixed by the angular ODE), but also of the full equation of motion, and this will also be true for the generalized CT inversion in Section~\ref{sec:SubextremalKerr}.
That is, they are conformal transformations that act within each superselection sector on the scalar field $\Phi_{\omega m}$, labeled by frequency $\omega$ and azimuthal number $m$ but not orbital number $\ell$ (which enters only after solving the angular problem), as
\begin{equation*}
    r-M \xrightarrow{\text{CT}} \tilde{r}-M = \frac{M^2+a^2-\frac{ma}{\omega}}{r-M} \,,\quad \Phi_{\omega m} \xrightarrow{\text{CT}} \tilde{\Phi}_{\omega m} = \frac{M}{r-M}\Phi_{\omega m} \,.
\end{equation*}
}
\be\label{eq:CTEKN}
	r-M \xrightarrow{\text{CT}} \tilde{r}-M = \frac{M^2+a^2-\frac{ma}{\omega}}{r-M} \,,\quad
	R_{\omega\ell m} \xrightarrow{\text{CT}} \tilde{R}_{\omega\ell m} = \frac{M}{r-M}R_{\omega\ell m} \,.
\ee
Hence, the EKN black hole also possesses an exact conformal symmetry of its wave equation.
As in ERN, this CT symmetry exchanges $r\to\infty$ with the radius $r=M$ of the (degenerate) horizon.
However, by contrast with the ERN case, where the radial inversion was independent of energy and angular momentum, these CT inversions depend explicitly on $(\omega,m)$ and therefore act differently within each superselection sector of fixed energy and angular momentum.
In other words, if the phase space of field configurations is foliated into leaves of fixed energy and angular momentum, then the CT inversion takes a different form across each leaf of the foliation.
This observation provides an intuitive explanation for why the symmetry cannot possibly manifest at the level of the metric itself, and therefore why the EKN black hole does not enjoy a discrete conformal isometry, unlike its ERN counterpart.

The existence of CT symmetry for extremal black holes\footnote{See, e.g., Refs.~\cite{Bianchi:2021yqs,Bianchi:2022wku,Cvetic:2020kwf,Cvetic:2021lss} for generalized CT inversions in the context of D$p$-branes and black holes of various dimensions in supergravity.} thus raises the question:
\begin{center}
    \textit{Could this symmetry be extended to generic (subextremal) black holes?} ($\star$)
\end{center}
A positive answer would represent a significant discovery in black hole physics.
This paper does not definitively address this question, but does take initial steps towards resolving it in the affirmative.
In particular, we show that the common objections to the possible existence of such an extension do not, in fact, apply.
In other words, we do not know of any obstruction that would forbid the extension of a CT-type symmetry to arbitrary spin.
In fact, in the ``static limit'' of low energy $\omega\to0$, we are able to explicitly construct a generalized CT symmetry that holds for all spins.
The relevant inversion takes a considerably more complicated form; this might explain why the full extension to all sectors $(\omega,m)$---should it indeed exist---has not yet been identified.\\

We know two pieces of ``lore'' that may prevent a generalization of CT symmetry beyond extremality.
The first one concerns the singularity structure of the radial ODE.
Let us briefly summarize this argument in the context of the Kerr black hole.

In general, the wave equation on the Kerr--(anti-)de Sitter black hole separates into a radial ODE that admits exactly four regular singular points.
These are located at the inner and outer horizons $r_\pm$, at the physical cosmological horizon $r_{\rm h}>0$, and at an unphysical cosmological horizon of negative radius.\footnote{Under a naive separation ansatz, one obtains a radial ODE with five singular points, including an additional one at infinity which, thanks to special properties of the equation, can be removed via a field redefinition and coordinate transformation (see, e.g., Appendix~G of Ref.~\cite{Berens:2025kkm} for details).}
The general second-order linear ODE with four singular points is the Heun equation, and its solutions define a special function HeunG that can be used to express the radial modes of the black hole \cite{Berens:2025kkm}.
In asymptotically flat Kerr, the two cosmological horizons are pushed out to infinity, which becomes an irregular singular point, and the radial modes may then be expressed in terms of a special function HeunC defined as a solution of the confluent Heun equation, in which two of the singular points have coalesced.
When this black hole becomes extremal, its two horizons merge ($r_\pm\to M$), leaving only two irregular singular points at horizon and infinity.
The CT inversion \eqref{eq:CTEKN} interchanges these two irregular singular points, leaving the overall singularity structure of the radial ODE unchanged.
The objection, therefore, is the following: for a subextremal Kerr black hole, with two regular singular points at $r_\pm$ and an irregular one at $r\to\infty$, exchanging infinity with either horizon does not preserve the singularity structure of the ODE (since an irregular point would be exchanged with a regular one).
Naively, this would seem to preclude an extension of CT symmetry to the subextremal regime.
However, since the CT inversions are conformal symmetries, they may modify the singularity structure of the ODE.
In Section~\ref{sec:LukewarmBH}, we illustrate this explicitly using the example of the so-called lukewarm black hole~\cite{Romans:1991nq}: an asymptotically de Sitter black hole in thermal equilibrium with its cosmological background.

A second, related ``lore'' is that the CT inversion exchanges the event horizon with infinity.
This idea is misleading as it becomes incorrect for a non-asymptotically-flat black hole, whose conformal boundary is not lightlike and thus can no longer be mapped to the event horizon.
We illustrate this in Section~\ref{sec:LukewarmBH} using once again the example of the lukewarm de Sitter black hole, for which the CT inversion is an exact conformal isometry that interchanges the event horizon with the cosmological one, rather than with infinity~\cite{Brannlund:2003gj}.
A better way to view the CT inversion is rather in terms of its action on null geodesics (see, e.g., Refs.~\cite{Bianchi:2021yqs,Bianchi:2022wku}).
A key property of the CT inversion is that it leaves invariant the photon sphere of bound null geodesics.\footnote{A geometric way of understanding the CT inversion is in terms of optical geometry~\cite{Brannlund:2003gj}.}
This property is natural given the lightlike nature of the photon sphere.
Interestingly, this observation turns out to be useful even in the case of a rotating Kerr black hole.
As a black hole spins up, the Schwarzschild photon sphere thickens into a ``photon shell'' containing a continuum of bound orbits.
In each superselection sector $(\omega,m)$, photons can asymptote to a unique bound orbit $\tilde{r}(m/\omega)$ and, as we show in Section~\ref{sec:PhotonSpheres}, the radial inversion within that sector preserves precisely that photon shell.

This paper is organized as follows.
In Section~\ref{sec:LukewarmBH}, we clarify why the two above arguments  against a generalized CT inversion are not valid, and illustrate this by using an explicit example: that of the static lukewarm black hole.
First, we recall that a CT inversion exists for this nonextremal geometry, and that it cannot be identified as a map between the horizon and infinity, but rather as an exchange of a pair of horizons (the event and cosmological horizons).
We then show that, in this geometry, the CT inversion does not preserve the singularity structure of the ODE governing probe fields.

Next, Section~\ref{sec:PhotonSpheres} examines CT inversions through the lens of their action on photon shells.
For the static cases of the ERN and the lukewarm de Sitter geometries, the photon spheres are fixed points of the CT inversions.
For the more intricate case of the EKN black hole, where CT inversions are phase-space-dependent, we show that the inversion preserves the photon shell in each superselection sector.
Then, Section~\ref{sec:Superradiance} argues that, for asymptotically flat spacetimes, fixed points of CT inversions are related to the phenomenon of superradiance.
We present a formula in which the superradiance coefficient controls the $(\omega,m)$-dependent part of the CT inversion, and prove its validity for a massless scalar field minimally coupled to gravity in the exrtremal Kerr--Newman geometry.
Finally, in Section~\ref{sec:SubextremalKerr}, we identify a new hidden CT symmetry for scalar perturbations of a Kerr black hole in the static and axisymmetric case $(\omega,m)=(0,0)$.
Taken together, these results suggest that CT symmetry might admit a broader generalization beyond extremality.

\section{CT symmetry of the static lukewarm de Sitter black hole}
\label{sec:LukewarmBH}

In this section, we consider two pieces of ``lore'' that would appear to obstruct the generalization of CT inversions beyond extremality.
These naive expectations are not in fact correct, as we show by exhibiting an explicit example that violates them both: the static lukewarm black hole~\cite{Romans:1991nq}.
This spacetime belongs to a class of asymptotically de Sitter (dS) Reissner-Nordstr\"om geometries, with line element
\be\label{eq:SphSymMetric}
	ds^2 = -\frac{\Delta_{r}}{r^2}\ed t^2 +\frac{r^2\ed r^2}{\Delta_{r}} +r^2\ed\Omega_2^2 \,,
\ee
where $\Delta_{r}$ is a discriminant function given in terms of the dS curvature radius $\ell$ by
\be\label{eq:LukewarmMetric}
    \Delta_{r}(r) = r^2\left(1-\frac{r^2}{\ell^2}\right)-2Mr+Q^2 \,.
\ee
The cosmological constant is $\Lambda=\frac{3}{\ell^2}$.
In general, the discriminant \eqref{eq:LukewarmMetric} admits four distinct real roots, corresponding to four globally defined null hypersurfaces: the cosmological horizon $r_{\rm c}$, the outer (event) horizon $r_{+}$, the inner (Cauchy) horizon $r_{-}$, and a negative-radius non-physical horizon $r_{\rm n}$, organized as $r_{\rm n} < 0\le r_{-}\le r_{+}\le r_{\rm c}$.
    
The ``lukewarm'' solution specifically refers to the black hole within this class whose temperature is the same as the dS thermal background, that is, whose event horizon is in equilibrium with the cosmological horizon~\cite{Romans:1991nq}.
This geometry has
\be
    Q^2 = M^2
    \quad\Longrightarrow\quad
    \Delta_{r} = -\frac{r^4}{\ell^2} + \left(r-M\right)^2 \,.
\ee
For this parameter choice, the surface gravity of the cosmological horizon exactly matches that of the event horizon.
As noted by Gibbons and Hawking \cite{Gibbons:1977mu}, such a configuration avoid problematic branch cuts in the thermodynamic description of asymptotically dS black holes.\footnote{Such branch cuts arise when one has two or more horizons with different surface gravities, for which one has to introduce separate Kruskal-type coordinate patches to penetrate each horizon.}
The static lukewarm black hole has four horizons at
\be\ba
	r_{\rm c} = \frac{\ell}{2}\left[1+\sqrt{1-\frac{4M}{\ell}}\right] &\,,\quad
	r_{+} = \frac{\ell}{2}\left[1-\sqrt{1-\frac{4M}{\ell}}\right] \,, \\
	r_{-} = \frac{\ell}{2}\left[-1+\sqrt{1+\frac{4M}{\ell}}\right] &\,,\quad
	r_{\rm n} = \frac{\ell}{2}\left[-1-\sqrt{1+\frac{4M}{\ell}}\right] \,,
\ea\ee
in terms of which the discriminant function takes the characteristic form
\be
	\Delta_{r} = -\frac{1}{\ell^2}\left(r-r_{\rm c}\right)\left(r-r_{+}\right)\left(r^2+\left(r_{\rm c}+r_{+}\right)r-r_{\rm c}r_{+}\right) \,.
\ee
Importantly, the static lukewarm geometry describes a \textit{nonextremal} black hole.

\subsection{Conformal inversion for a nonextremal horizon}

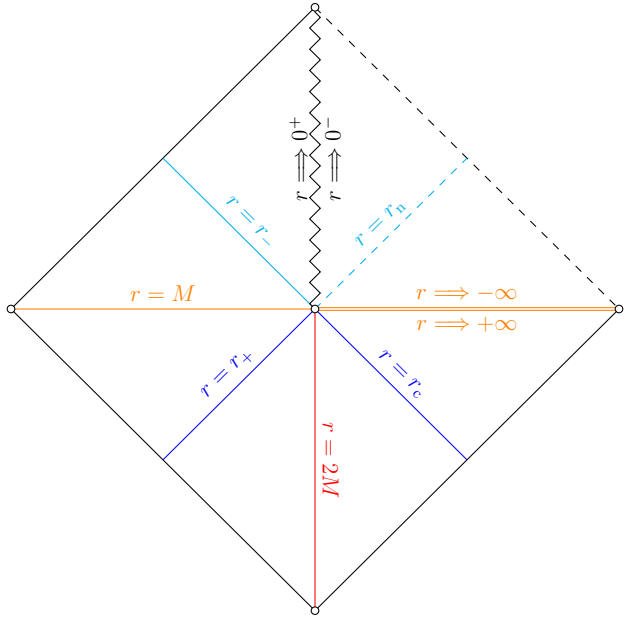
\begin{figure}
	\centering
	\begin{tikzpicture}
		\def\nL{4.0};
		\def\txtscale{0.18*\nL}
		\def\colrH{blue};
		\def\colrC{blue};
		\def\colrm{cyan};
		\def\colrn{cyan};
		\def\colrphInf{orange};
		\def\colrphz{orange};
		\def\colrphp{red};
		
		\draw (+\nL,0) -- (0,-\nL) -- (-\nL,0) -- (0,+\nL);
		\draw[dashed] (0,+\nL) -- (+\nL,0);
		
		\draw[\colrC] (0,0) -- node[scale=\txtscale, midway, above, sloped, \colrC]{$r=r_{\rm c}$} (0.5*\nL,-0.5*\nL);
		\draw[\colrH] (0,0) -- node[scale=\txtscale, midway, above, sloped, \colrH]{$r=r_{+}$} (-0.5*\nL,-0.5*\nL);
		\draw[\colrm] (0,0) -- node[scale=\txtscale, midway, above, sloped, \colrm]{$r=r_{-}$} (-0.5*\nL,+0.5*\nL);
		\draw[\colrn,dashed] (0,0) -- node[scale=\txtscale, midway, above, sloped, \colrn]{$r=r_{\rm n}$} (+0.5*\nL,+0.5*\nL);
		\draw[\colrphInf,double] (0,0) -- node[scale=\txtscale, midway, above, sloped, \colrphInf]{$r\Longrightarrow-\infty$} node[scale=\txtscale, midway, below, sloped, \colrphInf]{$r\Longrightarrow+\infty$} (+\nL,0);
		\draw[\colrphz] (0,0) -- node[scale=\txtscale, midway, above, sloped, \colrphInf]{$r=M$} (-\nL,0);
		\draw[\colrphp] (0,0) -- node[scale=\txtscale, midway, above, sloped, \colrphp]{$r=2M$} (0,-\nL);
		
		\draw [snake=zigzag, decorate, decoration={segment length=2*\nL, amplitude=0.5*\nL}] (0,0) -- node[scale=\txtscale, midway, above, sloped, black]{$r\Longrightarrow0^{+}$} (0,+\nL) node[scale=\txtscale, midway, below, sloped, black]{$r\Longrightarrow0^{-}$} (0,+\nL);;
		
		\node[draw, black, shape=circle, fill=white, minimum size = 0.1cm, inner sep=0pt] at (-\nL,0) () {};
		\node[draw, black, shape=circle, fill=white, minimum size = 0.1cm, inner sep=0pt] at (0,+\nL) () {};
		\node[draw, black, shape=circle, fill=white, minimum size = 0.1cm, inner sep=0pt] at (+\nL,0) () {};
		\node[draw, black, shape=circle, fill=white, minimum size = 0.1cm, inner sep=0pt] at (0,-\nL) () {};
		\node[draw, black, shape=circle, fill=white, minimum size = 0.1cm, inner sep=0pt] at (0,0) () {};
	\end{tikzpicture}
	\caption{Penrose diagram for the static lukewarm black hole showing surfaces of constant $r$ that are of particular interest (see Refs.~\cite{Mellor:1989gi,Mellor:1989wc,Runarsson:2012yh}), color-coded in CT-exchanged pairs.}
	\label{fig:LukewarmPenrose}
\end{figure}

It was already observed in Ref.~\cite{Brannlund:2003gj} (see also Ref.~\cite{Runarsson:2012yh}) that the static lukewarm black hole enjoys the CT conformal isometry
\be\label{eq:LukewarmCT}
	r-M \xrightarrow{\text{CT}} \tilde{r}-M = \frac{M^2}{r-M}
	\quad\Longrightarrow\quad
	r^{-2}ds^2 = \tilde{r}^{-2}d\tilde{s}^2 \,.
\ee
We now introduce the tortoise coordinate
\be\ba
	r_{\ast}(r) &= \frac{\ln\left|\frac{r_{+}}{r_{\rm c}}\frac{r-r_{\rm c}}{r-r_{+}}\right|}{2\left(r_{\rm c}-r_{+}\right)}+\frac{\ln\left|\frac{r_{\rm n}}{r_{-}}\frac{r-r_{-}}{r-r_{\rm n}}\right|}{2\left(r_{-}-r_{\rm n}\right)} \,,
\ea\ee
which maps the horizons to positive or negative infinity:
\be\ba
	r_{\ast} &\xrightarrow{r\Longrightarrow r_{\rm c}} -\infty \,,\qquad
	r_{\ast} &\xrightarrow{r\Longrightarrow r_{+}} +\infty \,, \\
	r_{\ast} &\xrightarrow{r\Longrightarrow r_{-}} -\infty \,,\qquad
	r_{\ast} &\xrightarrow{r\Longrightarrow r_{\rm n}} +\infty \,. \\
\ea\ee
In terms of this coordinate, the radial inversion \eqref{eq:LukewarmCT} is simply
\be
	r_{\ast} \xrightarrow{\text{CT}} \tilde{r}_{\ast} = -r_{\ast} \,.
\ee
In this form, the CT inversion manifestly exchanges pairs of horizons:
\be\ba
	r_{+} \,\text{(event horizon)} \quad  &\xleftrightarrow{\text{CT}} \quad r_{\rm c} \,\text{(cosmological horizon)} \,, \\
	r_{\rm n} \,\text{(negative root)} \quad &\xleftrightarrow{\text{CT}} \quad r_{-} \,\text{(inner horizon)} \,.
\ea\ee
Moreover, it exchanges $r=M$ with infinity, while keeping the $r=2M$ surface fixed:
\be\ba
	r=M \quad  &\xleftrightarrow{\text{CT}} \quad r=\infty \,, \\
	r=2M \quad &\xleftrightarrow{\text{CT}} \quad r=2M \,.
\ea\ee
As we will see in Section~\ref{subsec:LukewarmPhotonSphere}, the $r=2M$ surface is precisely the photon sphere of unstably bound photon orbits that the static lukewarm black hole is equipped with.
We illustrate the action of CT inversions on the static lukewarm black hole in Fig.~\ref{fig:LukewarmPenrose}.

The static lukewarm black hole provides an explicit example where the naive expectations discussed in the Introduction fail.
First of all, since the lukewarm black hole is \textit{nonextremal}, it proves that CT symmetry is not necessarily limited to extremal black holes.
In fact, the static lukewarm geometry only describes a black hole if $\ell\ge4M$, but the CT inversion \eqref{eq:LukewarmCT} is a discrete conformal isometry of the $Q^2=M^2$ RN-dS solution for \textit{any} value of $\ell$, including in the range $\ell<4M$ for which the spacetime becomes a naked singularity.
More generally, the CT inversion \eqref{eq:LukewarmCT} is a conformal isometry for any sign of the cosmological constant, persisting when
\be\label{eq:LukewarmK}
	\Delta_{r}(r) = -k\frac{r^4}{\ell^2} + \left(r-M\right)^2 \,.
\ee
In this case, the Einstein-Maxwell geometry \eqref{eq:SphSymMetric} is a static electrovacuum solution with cosmological constant $\Lambda=\frac{3k}{\ell^2}$.
The parameter $k$ is thus equal to $+1$, $-1$, or $0$ according to whether the RN spacetime with charge $Q^2=M^2$ is asymptotically dS, AdS, flat, respectively.
This spacetime describes a black hole only for $k=+1$ and $\ell\ge4M$ (static lukewarm black hole) or $k=0$ (ERN black hole); otherwise, it contains a naked singularity (in which case the CT-pairs of null hypersurfaces acquire complex-valued radii and become unphysical).

Second, the CT inversion \eqref{eq:LukewarmCT} does \textit{not} exchange the event horizon with infinity, but rather with the cosmological horizon.
This should not come as a surprise, since the CT inversion is a conformal isometry and hence must exchange null hypersurfaces, whereas the conformal boundary is spacelike for asymptotically dS spacetimes, rather than null as in the $k=0$ (ERN) case.

\subsection{CT inversions need not preserve the ODE singularity structure}

In addition, the static lukewarm black hole shows that a CT symmetry need \textit{not} preserve the singularity structure of the equations of motion for probe fields.
Consider, for example, the wave equation for a conformally coupled real scalar field:
\be
	\left(\nabla^2 - \frac{R}{6}\right)\Phi = 0 \,.
\ee
The conformal coupling to gravity ensures that this equation is invariant under all conformal isometries of the background geometry.
In the static lukewarm black hole spacetime, for which $R=4\Lambda=\frac{12}{\ell^2}$, this equation of motion separates into the following radial ODE for the monochromatic spherical harmonic modes $\Phi_{\omega\ell m}$ of frequency $\omega$, azimuthal number $m$ and orbital number $\ell$:
\be\label{eq:WaveEqSingularities1}
	\left[\partial_{r}^2+p_1(r)\partial_{r}+p_0(r)\right]\Phi_{\omega\ell m} = 0 \,,
\ee
with
\be\label{eq:WaveEqSingularities2}
	p_1(r) = \frac{\Delta_{r}^{\prime}(r)}{\Delta_{r}(r)} \,,\quad
	p_0(r) = \frac{1}{\Delta_{r}(r)}\left[\frac{\omega^2r^4}{\Delta_{r}(r)} -\ell\left(\ell+1\right)-2\frac{r^2}{\ell^2}\right] \,.
\ee

Let us see what happens to the singular points of this radial ODE under the CT inversion \eqref{eq:LukewarmCT}.
First, we consider the asymptotically flat ERN spacetime, which is recovered in the $\ell\Longrightarrow\infty$ limit of the static lukewarm black hole.
A direct calculation shows that the extremal event horizon at $r=M$ is an irregular singular point of the radial ODE, while $r=2M$ is an ordinary point.
As for the $r\Longrightarrow\infty$ point, we first need to perform the $u=\frac{1}{r}$ transformation of the independent variable and bring the resulting ODE into the canonical form $\left[\partial_{u}^2 + \tilde{p}_1\left(u\right)\partial_{u} + \tilde{p}_0\left(u\right)\right]\Phi_{\omega\ell m} = 0$.
In so doing, we find that $r\Longrightarrow\infty$ is also an irregular singular point.
This is consistent with the naive ``no-go'' expectation that the CT inversion always maps a singular point of the ODE onto another singular point of the same rank.

Next, consider the asymptotically (A)dS geometry of the static lukewarm spacetime with finite curvature radius $\ell$.
Repeating the same procedure, one can check that $r\Longrightarrow\infty$ is now a regular singular point, while its CT-related point at $r=M$ is, in fact, an ordinary point of the radial ODE.
Thus, CT symmetry need not preserve the singularity structure of the radial ODE.
For convenience, we summarize with a table the singularity structure of the ODE near the CT-related points $r=M$ and $r\Longrightarrow\infty$, as well as near its fixed point $r=2M$, for each type of asymptotic behavior:
\begin{table}[H]
	\centering
	\begin{tabular}{|c|c|c|}
		\hline
		Point of Interest & $\begin{matrix*} \text{Singularity of ODE for} \\ \text{asymptotically flat spacetime} \end{matrix*}$ & $\begin{matrix*} \text{Singularity of ODE for} \\ \text{asymptotically (A)dS spacetime} \end{matrix*}$ \\
		\hline
		\hline
		$r=M$ & Irregular singular point & Ordinary point \\
		$r\Longrightarrow\infty$ & Irregular singular point & Regular singular point \\
		$r=2M$ & Ordinary point & Ordinary point \\
		\hline
	\end{tabular}
	\caption{Singularity structure of the radial ODE \eqref{eq:WaveEqSingularities1}--\eqref{eq:WaveEqSingularities2} for scalar-field modes in the asymptotically flat or (A)dS geometries \eqref{eq:SphSymMetric} with discriminant \eqref{eq:LukewarmK}.}
	\label{tbl:GenCTODESingularityPhi}
\end{table}

This explicit counterexample shows that the naive no-go argument against the existence of a CT inversion beyond extremality does not apply.
This is because CT inversions are \textit{conformal} symmetries of the equations of motion.
As a result, while it is true that the singularity structure of an ODE must be preserved under isometric changes of the independent variables, this need not be true once we allow for changes of the dependent variables as well, such as in conformal transformations.
This interplay between exact and conformal isometries at the level of an ODE can be made explicit for the conformally coupled scalar field perturbation of the static lukewarm black hole by considering the redefined scalar field
\be
	\varphi = \sqrt{\frac{M}{r-M}}\,\Phi \,.
\ee
This redefinition makes $\varphi$ conformally weightless under CT inversions and, hence, the CT inversions are now true isometries of the radial ODE.
If one works with these variables, then the resulting radial ODE has a singularity structure that is \textit{different} than the initial ODE we noted above, and this different singularity structure \textit{is} exactly preserved under CT inversions.
More explicitly, the conformal wave equation \eqref{eq:WaveEqSingularities1}--\eqref{eq:WaveEqSingularities2} for the separable field $\varphi_{\omega\ell m}$ becomes
\be\label{eq:WaveEqSingularitiesRed1}
	\left[\partial_{r}^2 + P_1(r)\partial_{r} + P_0(r)\right]\varphi_{\omega\ell m} = 0 \,,
\ee
with
\be\label{eq:WaveEqSingularitiesRed2}
	P_1(r) = p_1(r) + \frac{1}{r-M} \,,\quad
	P_0(r) = p_0(r) + \frac{p_1(r)}{2\left(r-M\right)} -\frac{1}{4\left(r-M\right)^2} \,.
\ee
As a result, only the singularity structure near the point at $r=M$ is affected by this redefinition of the dependent variable.
Indeed, while it is an ordinary point for the ODE of the field $\Phi_{\omega\ell m}$, it is a regular singular point for the ODE of the field $\varphi_{\omega\ell m}$.
The full singularity structures of the equations of motion for the conformally coupled scalar perturbations $\Phi$ of the static lukewarm black hole and for the redefined field $\varphi=\sqrt{\frac{M}{r-M}}\,\Phi$ are summarized in Fig.~\ref{fig:LukewarmODESingularityPhiVarphi} below.

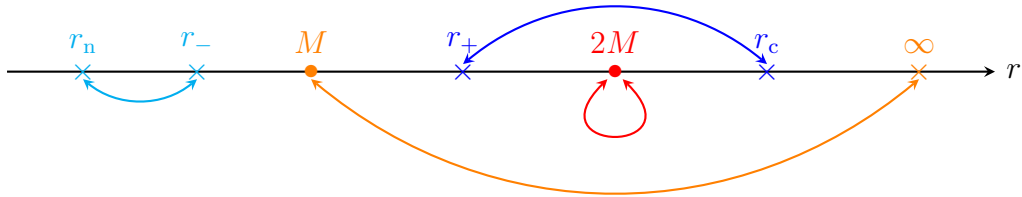
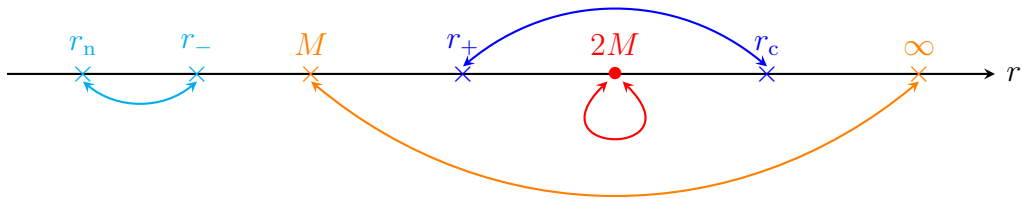
\begin{figure}[ht]
\centering
\begin{subfigure}[h]{1.0\textwidth}
\centering
\begin{tikzpicture}[>=stealth]
    \def\nL{1.0};
    \draw[thick, ->] (0,0) -- (13*\nL,0) node[right] {$r$};

    \coordinate (rn) at (1.0*\nL,0.0);
    \coordinate (rminus) at (2.5*\nL,0.0);
    \coordinate (rm) at (4.0*\nL,0.0);
    \coordinate (rh) at (6.0*\nL,0.0);
    \coordinate (r2m) at (8.0*\nL,0.0);
    \coordinate (rc) at (10.0*\nL,0.0);
    \coordinate (inf) at (12*\nL,0.0);

    \node at (rn) {\textcolor{cyan}{$\times$}};
    \node[above=3pt] at (rn) {\textcolor{cyan}{$r_{\rm n}$}};

    \node at (rminus) {\textcolor{cyan}{$\times$}};
    \node[above=3pt] at (rminus) {\textcolor{cyan}{$r_{-}$}};

    \node at (rm) {\textcolor{orange}{$\bullet$}};
    \node[above=3pt] at (rm) {\textcolor{orange}{$M$}};

    \node at (rh) {\textcolor{blue}{$\times$}};
    \node[above=3pt] at (rh) {\textcolor{blue}{$r_{+}$}};

    \node at (r2m) {\textcolor{red}{$\bullet$}};
    \node[above=3pt] at (r2m) {\textcolor{red}{$2M$}};

    \node at (rc) {\textcolor{blue}{$\times$}};
    \node[above=3pt] at (rc) {\textcolor{blue}{$r_{\rm c}$}};

    \node at (inf) {\textcolor{orange}{$\times$}};
    \node[above=3pt] at (inf) {\textcolor{orange}{$\infty$}};

    \draw[<->, thick, cyan, out=-40, in=220] ($(rn)+(0.0*\nL,-0.1*\nL)$) to ($(rminus)+(0.0*\nL,-0.1*\nL)$);
    \draw[<->, thick, orange, out=-40, in=220] ($(rm)+(0.0*\nL,-0.1*\nL)$) to ($(inf)+(0.0*\nL,-0.1*\nL)$);
    \draw[<->, thick, blue, out=40, in=-220] ($(rh)+(0.0*\nL,+0.1*\nL)$) to ($(rc)+(0.0*\nL,+0.1*\nL)$);
    \draw[<->, thick, red, out=220, in=-40, looseness=2] ($(r2m)+(-0.1*\nL,-0.1*\nL)$)
      .. controls +(-1.0*\nL,-1.0*\nL) and +(+1.0*\nL,-1.0*\nL) ..
    ($(r2m)+(+0.1*\nL,-0.1*\nL)$);
\end{tikzpicture}
\caption{For $\Phi$.}
\label{fig:LukewarmODESingularityPhi}
\end{subfigure}
\begin{subfigure}[h]{1.0\textwidth}
\centering
\begin{tikzpicture}[>=stealth]
    \def\nL{1.0};
    \draw[thick, ->] (0,0) -- (13*\nL,0) node[right] {$r$};

    \coordinate (rn) at (1.0*\nL,0.0);
    \coordinate (rminus) at (2.5*\nL,0.0);
    \coordinate (rm) at (4.0*\nL,0.0);
    \coordinate (rh) at (6.0*\nL,0.0);
    \coordinate (r2m) at (8.0*\nL,0.0);
    \coordinate (rc) at (10.0*\nL,0.0);
    \coordinate (inf) at (12*\nL,0.0);

    \node at (rn) {\textcolor{cyan}{$\times$}};
    \node[above=3pt] at (rn) {\textcolor{cyan}{$r_{\rm n}$}};

    \node at (rminus) {\textcolor{cyan}{$\times$}};
    \node[above=3pt] at (rminus) {\textcolor{cyan}{$r_{-}$}};

    \node at (rm) {\textcolor{orange}{$\times$}};
    \node[above=3pt] at (rm) {\textcolor{orange}{$M$}};

    \node at (rh) {\textcolor{blue}{$\times$}};
    \node[above=3pt] at (rh) {\textcolor{blue}{$r_{+}$}};

    \node at (r2m) {\textcolor{red}{$\bullet$}};
    \node[above=3pt] at (r2m) {\textcolor{red}{$2M$}};

    \node at (rc) {\textcolor{blue}{$\times$}};
    \node[above=3pt] at (rc) {\textcolor{blue}{$r_{\rm c}$}};

    \node at (inf) {\textcolor{orange}{$\times$}};
    \node[above=3pt] at (inf) {\textcolor{orange}{$\infty$}};

    \draw[<->, thick, cyan, out=-40, in=220] ($(rn)+(0.0*\nL,-0.1*\nL)$) to ($(rminus)+(0.0*\nL,-0.1*\nL)$);
    \draw[<->, thick, orange, out=-40, in=220] ($(rm)+(0.0*\nL,-0.1*\nL)$) to ($(inf)+(0.0*\nL,-0.1*\nL)$);
    \draw[<->, thick, blue, out=40, in=-220] ($(rh)+(0.0*\nL,+0.1*\nL)$) to ($(rc)+(0.0*\nL,+0.1*\nL)$);
    \draw[<->, thick, red, out=220, in=-40, looseness=2] ($(r2m)+(-0.1*\nL,-0.1*\nL)$)
      .. controls +(-1.0*\nL,-1.0*\nL) and +(+1.0*\nL,-1.0*\nL) ..
    ($(r2m)+(+0.1*\nL,-0.1*\nL)$);
\end{tikzpicture}
\caption{For $\varphi$.}
\label{fig:LukewarmODESingularityVarphi}
\end{subfigure}
\caption{Singularity structure of the wave equation for conformally coupled scalar perturbations of the static lukewarm black hole.
	Top: Fig.~\ref{fig:LukewarmODESingularityPhi} shows this structure for the wave operator \eqref{eq:WaveEqSingularities1}--\eqref{eq:WaveEqSingularities2} describing the propagation of the canonical scalar field $\Phi$.
	Bottom: Fig.~\ref{fig:LukewarmODESingularityVarphi} shows this structure for the wave operator \eqref{eq:WaveEqSingularitiesRed1}--\eqref{eq:WaveEqSingularitiesRed2} governing the field $\varphi=\sqrt{\frac{M}{r-M}}\,\Phi$ that is conformally weightless under CT inversion.
	A ``$\times$'' or ``$\bullet$'' represents a regular singular point or an ordinary point of the ODE, respectively.
	The arrows connect radial points that are related by a CT inversion.}
\label{fig:LukewarmODESingularityPhiVarphi}
\end{figure}

In summary, we have exhibited an explicit example invalidating the conventional ``no-go'' arguments presented in the Introduction against the possible generalization of CT symmetry beyond extremality.

\section{CT inversions preserve photon spheres}
\label{sec:PhotonSpheres}

Having seen that the usual ``no-go'' lore does not necessarily obstruct CT symmetry beyond extremality, in this section we ask what a more general CT inversion should look like.
A useful guide is to step back from the field equations and instead examine the geometric imprint shared by the examples that we already understand.

So far, we have emphasized CT inversions as conformal symmetries of the probe-field equations of motion.
Now, we recast the discussion in terms of null geodesics, and in particular the structure of spherical photon orbits (photon spheres and photon shells).
This viewpoint has appeared in related contexts; see, e.g., Refs.~\cite{Bianchi:2021yqs,Bianchi:2022wku}.

As reviewed in Appendix~\ref{app:GeodesicsKNLambda}, the null geodesic equations in the Kerr--Newman and Kerr--Newman--(A)dS spacetimes are completely integrable.
The stationarity and axisymmetry of these spacetimes imply that their geodesics have a conserved energy $E$ and a conserved axial angular momentum $L_z$, respectively.
In addition, they carry a third independent constant of motion: the Carter constant $\mathcal{C}$, whose existence follows from a hidden symmetry generated by a rank-$2$ Killing tensor~\cite{Carter:1968rr}.
These integrals lead to the separable first-order system \eqref{eq:KNLambdaGeodesicsEqs}.
For our purposes, the key piece is the radial equation, which can be written as 
\be
	\frac{\Sigma}{E}\dot{r} = \pm_{r}\sqrt{\mathcal{R}(r)} \,,
\ee
where $\Sigma=r^2+a^2\cos^2\theta$ and $\pm_r\equiv \mathrm{sign}(\dot r)$.
For electrically neutral null geodesics, the effective radial potential takes the form
\be
	\mathcal{R}(r) = \left(r^2+a^2-ab\right)^2 -\Delta_{r}\left(\eta+\left(a-b\right)^2\right) \,,
\ee
controlled by the impact parameter $b$ and the energy-rescaled Carter constant $\eta$,
\be
	b = \frac{L_{z}}{E}
    \quad\text{and}\quad
    \eta = \frac{\mathcal{C}}{E^2} \,.
\ee
Here, $\Delta_r$ denotes the usual radial discriminant for a  Kerr--Newman geometry, which for asymptotically dS ($k=+1$), AdS ($k=-1$), or flat ($k=0$) spacetimes is
\be
	\Delta_{r} = \left(r^2+a^2\right)\left(1-k\frac{r^2}{\ell^2}\right) -2Mr +Q^2 \,.
\ee

Spherical photon orbits correspond to bound null geodesics at constant radius $r=r_{\rm ph}$. Equivalently, they are double roots of the radial potential~\cite{Carter:1968rr,Wilkins:1972rs,Teo:2003ltt,Teo:2020sey},
\be\label{eq:PhotonShellsConditions}
	\mathcal{R}\left(r_{\rm ph}\right) = 0 \,,\quad
    \mathcal{R}^{\prime}\left(r_{\rm ph}\right) = 0 \,.
\ee
In a non-rotating spacetime, spherical symmetry dictates that these spherical photon orbits span the entire surface of a two-dimensional ``photon sphere''.
In general, there can be multiple photon spheres, with at least one unstable photon sphere lying in the black hole exterior~\cite{Claudel:2000yi,Cunha:2020azh,Ghosh:2021txu,Qiao:2022hfv,Qiao:2022jlu,Cunha:2022nyw,Qiao:2024qsf}.

\subsection{ERN black hole}

To study the extremal Reissner--Nordstr\"om (ERN) geometry, we set $a=0$, so that $\Delta_r=(r-M)^2$ and the spacetime is spherically symmetric.
In this case, the conditions \eqref{eq:PhotonShellsConditions} admit a single spherical photon orbit in the exterior region, located at
\be
	r_{\rm ph}=2M\,.
\ee
Evaluating the second derivative of the radial potential at this point shows
that the orbit is unstable, i.e.\ small radial perturbations drive the null
geodesic either inward toward the horizon or outward toward infinity.

The action of the CT inversion \eqref{eq:ERNCT} on this structure is particularly transparent: it exchanges the horizon at $r=M$ with null infinity,
\be
	r=M \quad \xleftrightarrow{\text{CT}} \quad r\to\infty\,,
\ee
while leaving the unstable photon sphere fixed,
\be
	r_{\rm ph}=2M \quad \xrightarrow{\text{CT}} \quad 2M\,.
\ee
Thus, in ERN the distinguished radius singled out by the CT inversion is precisely the photon-sphere radius.

This pattern---CT radial inversions move horizons and infinity while fixing the photon sphere/shell---will reappear in each of the examples we consider below.

\subsection{Static lukewarm geometry}
\label{subsec:LukewarmPhotonSphere}

We next turn to the example of the static lukewarm geometry, for which $a=0$ and $\Delta_{r}=-k\frac{r^4}{\ell^2}+\left(r-M\right)^2$.
Although the global causal structure depends on $k$ and on the curvature scale
$\ell$, the photon-sphere radius does not: solving
\eqref{eq:PhotonShellsConditions} again yields
\be
	r_{\rm ph}=2M\,,
\ee
with the value of $\mathcal{R}''(r_{\rm ph})$ showing that this photon sphere is unstable.
In other words, the lukewarm deformation leaves the (outer) photon sphere at the same location as in ERN.

The associated CT inversion \eqref{eq:LukewarmCT} therefore exhibits the same geometric pattern: it exchanges the CT-paired radii $r=M$ and $r\to\infty$ while fixing the unstable photon sphere at $r=2M$.
The important difference from ERN is that, for $k\neq 0$, the hypersurfaces $r=M$ and $r\to\infty$ are no longer null.
For example, in the asymptotically dS case ($k=+1$) one finds the ordering
\be
    r_{\rm n}< 0< r_{-}< M< r_{+}< 2M< r_{\rm c}< \infty \,,
\ee
so that both $r=M$ and $r\to\infty$ are spacelike, whereas the CT map exchanges the event and cosmological horizons, as discussed in Section~\ref{sec:LukewarmBH}.

From the geodesic perspective, the reason the photon sphere is preserved is simple: the CT map rescales the radial potential by an overall conformal factor.
Concretely, under the transformation \eqref{eq:LukewarmCT}, one finds that
\be
	\mathcal{R}(r) \xrightarrow{\text{CT}} \mathcal{R}\left(M+\frac{M^2}{r-M}\right) = \frac{M^4}{\left(r-M\right)^4}\mathcal{R}(r) \,.
\ee
Since the zeros of $\mathcal{R}$ (and the condition of being a double zero) are unchanged by an overall nonvanishing prefactor, the set of spherical photon orbits is invariant, and in particular the fixed point $r=2M$ is preserved.
We will now see how this statement generalizes to the rotating case, where a single photon sphere thickens into a continuous photon shell.

\subsection{EKN black hole}

When the black hole rotates, spherical symmetry is lost and a single photon sphere generically ``thickens'' into a photon shell: a continuous family of spherical photon orbits labeled by phase-space data \cite{Johnson:2019ljv}.\footnote{That the shell has finite thickness is not automatic; it follows from combining the existence conditions for spherical orbits with the reality constraints from the angular motion; see Appendix~\ref{app:GeodesicsKNLambda}.}
We illustrate this phenomenon for the extremal Kerr--Newman (EKN) geometry, for which $\Delta_{r}=\left(r-M\right)^2$.
In this case, for each allowed value of the impact parameter $b$, there is a
unique spherical photon orbit in the black hole exterior at
\be
	r_{\rm ph}(b) = M+\sqrt{M^2+a^2-ab} \,,
\ee
which is unstable; the union of these radii over the allowed range of $b$ forms the (outer) photon shell.

A key point for our purposes is that the CT inversions act naturally at the level of the radial potential.
In direct analogy with the field-theory discussion, the EKN CT map is phase-space dependent: it depends on the conserved quantities through $b=L_z/E$.
Explicitly, the effective radial potential associated with null geodesics is conformally preserved under the CT inversion\footnote{A similar phase-space-dependent CT inversion exists for electrically charged null geodesics, with specific charge $\mathfrak{q}$ and effective radial potential $\mathcal{R}(r)=\left(r^2+a^2-ab-\mathfrak{q}r\right)^2-\Delta_{r}\left(\eta+\left(a-b\right)^2\right)$.
More explicitly, for the EKN geometry with $\Delta_{r}=\left(r-M\right)^2$,
\begin{equation*}
    \begin{split}
        {}& r-M \xrightarrow{\text{CT}} \tilde{r}-M = \frac{M^2+a^2-ab-\mathfrak{q}M}{r-M} \,, \\
        \quad\Longrightarrow\quad
        & \mathcal{R}(r) \xrightarrow{\text{CT}} \mathcal{R}\left(\tilde{r}\right) = \frac{\left(M^2+a^2-ab-\mathfrak{q}r\right)^2}{\left(r-M\right)^4}\mathcal{R}(r) \,.
    \end{split}
\end{equation*}}
\be\ba
	{}& r-M \xrightarrow{\text{CT}} \tilde{r}-M = \frac{M^2+a^2-ab}{r-M} \,, \\
	\quad\Longrightarrow\quad& \mathcal{R}(r) \xrightarrow{\text{CT}} \mathcal{R}\left(\tilde{r}\right) = \frac{\left(M^2+a^2-ab\right)^2}{\left(r-M\right)^4}\mathcal{R}(r) \,.
\ea\ee
Thus the CT inversion preserves the zero locus of $\mathcal{R}$---and hence the spherical-orbit conditions \eqref{eq:PhotonShellsConditions}---within each superselection sector that is specified by $(E,L_z,\mathcal{C})$, or equivalently by $(b,\eta)$.

In particular, each spherical photon orbit is a fixed point of the corresponding CT inversion:
\be
	r_{\rm ph}(b) \xrightarrow{\text{CT}} \tilde{r}_{\text{ph}}(b) = r_{\rm ph}(b) \,.
\ee
Equivalently, for fixed $b$ the inversion acts as a reflection of the radial problem about the radius $r=r_{\rm ph}(b)$, in the sense that it exchanges the two asymptotic ends of the effective one-dimensional motion while leaving the (unstable) orbit itself invariant.

We thus arrive at a unified geometric statement shared by the static and rotating examples: CT inversions preserve the relevant photon region (a photon sphere in the spherically symmetric case, or a photon shell in the rotating case), and the fixed points of the inversion coincide with the corresponding spherical photon orbits.
It is therefore natural to expect this property to persist in any generalized CT inversions.

\section{Fixed points of CT inversions and superradiance}
\label{sec:Superradiance}

We will now relate the fixed points of CT inversions to another physical ingredient that naturally depends on the superselection data $(\omega,m)$: superradiance~\cite{Penrose:1969pc,Misner:1972kx,ZelDovich:1971,ZelDovich:1972,Press:1972zz,Starobinsky:1973aij,Starobinskil:1974nkd}.
This connection will allow us to factor all the known CT inversions into a purely geometric part (which reflects the tortoise coordinate) and a phase-space-dependent part (which is controlled by the superradiance coefficient), in a way that precisely reflects their actions on both null surfaces and spherical photon orbits.

For the separable black holes of interest, the scattering problem reduces (after mode decomposition) to a one-dimensional radial equation of Schr\"odinger type.
The superradiance coefficient $z$ appears in the flux-balance relation
\be
	\left|\mathcal{R}\right|^2 = \left|\mathcal{I}\right|^2 -z\left|\mathcal{T}\right|^2 \,,
\ee
between the reflection, transmission, and incidence amplitudes $\mathcal{R}$, $\mathcal{T}$ and $\mathcal{I}$.
The phenomenon of superradiant amplification occurs when $z<0$, i.e., when the reflected flux exceeds the incident one.
Equivalently, the amplification factor
\be
	Z = \frac{\left|\mathcal{R}\right|^2}{\left|\mathcal{I}\right|^2}-1 = -z\frac{\left|\mathcal{T}\right|^2}{\left|\mathcal{I}\right|^2}
\ee
is positive precisely when $z<0$, so that the reflected field has larger amplitude than the incident one~\cite{Starobinsky:1973aij,Starobinskil:1974nkd}.

The coefficient $z$ admits a simple interpretation as a ratio of asymptotic radial momenta.
Concretely, for a completely integrable stationary and axisymmetric black hole (such as the Kerr--Newman), after separation one may write the radial ODE as
\be\label{eq:RadialODEGeneral}
	\left[\frac{d^2}{dr_{\ast}^2} +\omega^2-\mathcal{V}_{\omega\ell m}(r)\right]\psi_{\omega\ell m}(r) = 0 \,,
\ee
where $r_{\ast}(r)$ is the tortoise coordinate, $\mathcal{V}_{\omega\ell m}$ is a real effective potential, and $\psi_{\omega\ell m}(r)$ is the radiative radial wavefunction associated with the scattered field.

We now specialize to asymptotically flat spacetimes,
for which $\mathcal{V}_{\omega\ell m}(r)=\mathcal{O}(r^{-1})$ as $r\to\infty$.
In that case, the radial momentum at null infinity obeys the simple characteristic dispersion relation $k_{\infty}\left(\omega,m\right)=\omega$, while near the event horizon $r=r_+$ (where $r_\ast\to-\infty$) the solutions
approach plane waves with momentum $k_+(\omega,m)$.
Imposing purely ingoing boundary conditions at the horizon and a superposition of incoming and outgoing waves at infinity gives
\be
	\psi_{\omega \ell m} \sim
	\begin{cases}
		\mathcal{T}e^{-ik_{+}\left(\omega,m\right)r_{\ast}} & \text{as $r_{\ast}\Longrightarrow-\infty$} \,, \\
		\mathcal{I}e^{-i\omega r_{\ast}} + \mathcal{R}e^{i\omega r_{\ast}} & \text{as $r_{\ast}\Longrightarrow+\infty$} \,.
	\end{cases}
\ee
Radial independence of the Wronskian between $\psi_{\omega\ell m}$ and its linearly-independent complex conjugate counterpart $\psi_{\omega\ell m}^{\ast}$ then implies that
\be\ba
	{}&\left(\psi_{\omega\ell m}\psi_{\omega\ell m}^{\ast\prime}-\psi_{\omega\ell m}^{\ast}\psi_{\omega\ell m}^{\prime}\right)\big|_{r_{\ast}\Longrightarrow+\infty} = \left(\psi_{\omega\ell m}\psi_{\omega\ell m}^{\ast\prime}-\psi_{\omega\ell m}^{\ast}\psi_{\omega\ell m}^{\prime}\right)\big|_{r_{\ast}\Longrightarrow-\infty} \,, \\
	&\Longrightarrow\quad \left|\mathcal{R}\right|^2 = \left|\mathcal{I}\right|^2 -\frac{k_{+}\left(\omega,m\right)}{\omega}\left|\mathcal{T}\right|^2 \,,
\ea\ee
showing that the superradiance coefficient is a ratio of asymptotic radial momenta,
\be
	z\left(\omega,m\right) = \frac{k_{+}\left(\omega,m\right)}{\omega} = \frac{k_{+}\left(\omega,m\right)}{k_{\infty}} \,.
\ee
In general, $z$ depends not only on $(\omega,m)$ but also on the quantum numbers specifying how the probe couples to the background (for instance, its electric charge).

We will now show that, in every known asymptotically flat instance where a CT inversion acts as a conformal symmetry of the separated field equations, the radial map can be written in the factorized form
\be\label{eq:CTgeneralExtremal}
	r - r_{\rm ph}^{\left(0\right)} \xrightarrow{\text{CT}} \tilde{r} - r_{\rm ph}^{\left(0\right)} = \frac{r_1^2}{r-r_{\rm ph}^{\left(0\right)}}z\left(\omega,m\right) \,.
\ee
This expression isolates two conceptually distinct ingredients:
\begin{itemize}
	\item $r_{\rm ph}^{(0)}$ denotes the distinguished spherical photon orbit that is CT-exchanged with infinity.
    For all the examples of interest in this work, $r_{\rm ph}^{\left(0\right)}=M$.
	\item The constant $r_1^2$ is a geometric length-squared scale fixed by the	requirement that the \emph{geometric} part of the map reflects the tortoise 	coordinate,
	\be
		r_{\ast}\left(r_{\rm ph}^{\left(0\right)}+\frac{r_1^2}{r-r_{\rm ph}^{\left(0\right)}}\right)=-r_{\ast}(r) \,.
	\ee
	\item The remaining dependence on the mode labels is entirely captured by the superradiance coefficient $z(\omega,m)$, i.e., by the ratio of the asymptotic radial momenta.
\end{itemize}
In this way, the CT inversion simultaneously encodes both the reflection of the one-dimensional scattering problem in tortoise coordinate, as well as the phase-space dependence required once the horizon and infinity ``see'' different
effective frequencies.

To justify the factorization \eqref{eq:CTgeneralExtremal} explicitly, consider a massless complex scalar $\Psi$ of electric charge $q$ minimally coupled to an asymptotically flat Kerr--Newman (KN) background.
The equation of motion for $\Psi$ is
\be
	D_\mu D^\mu\Psi=0\,,\qquad D_\mu=\nabla_\mu-iqA_\mu\,,
\ee
with $A_\mu$ the background Maxwell potential given in Eq.~\eqref{eq:KNLambdaMetric1}.
Decomposing into monochromatic spheroidal modes of definite frequency $\omega$ and azimuthal number $m$, $\Psi_{\omega\ell m}=e^{-i\omega t+im\phi}R_{\omega\ell m}(r)S_{\omega\ell m}\left(\theta\right)$, the radial mode obeys the ODE
\be
	\left[\frac{d}{dr}\left(r-M\right)^2\frac{d}{dr} +\frac{\left(r^2+a^2\right)^2}{\left(r-M\right)^2}\left(\omega-\frac{ma+qQr}{r^2+a^2}\right)^2 -A_{\omega\ell m}\right]R_{\omega\ell m} = 0 \,,
\ee
where $A_{\omega\ell m}$ is the separation constant fixed by the angular eigenvalue problem.
Studying the asymptotics near the degenerate event horizon and near null infinity, one then identifies the following asymptotic radial momenta\footnote{This radial wavefunction is related to the radiative one $\psi_{\omega\ell m}$ in Eq.~\eqref{eq:RadialODEGeneral} by $R_{\omega\ell m}=\frac{\psi_{\omega\ell m}}{\sqrt{r^2+a^2}}$.}
\be
	k_{+} = \omega -m\Omega_{\rm H} -q\Phi_{\rm H} \,,\quad
    k_{\infty} = \omega \,,
\ee
where
\be
	\Omega_{\rm H} = \frac{a}{M^2+a^2}
    \quad\text{and}\quad
    \Phi_{\rm H} = \frac{QM}{M^2+a^2}
\ee
are the angular velocity and electrostatic potential of the horizon, respectively.
The superradiance coefficient is then the ratio of these radial momenta, and the proposed CT inversions of Eq.~\eqref{eq:CTgeneralExtremal} read
\be
	r-M \xrightarrow{\text{CT}} \tilde{r}-M = \frac{M^2+a^2}{r-M}z \,,\quad
    z = \frac{k_{+}}{k_{\infty}} = \frac{\omega -m\Omega_{\rm H} -q\Phi_{\rm H}}{\omega} \,.
\ee
where we used the formula for the tortoise coordinate of the EKN geometry,\footnote{Here, we fix without loss of generality the integration constant such that $r_{\ast}\left(r=M+\sqrt{M^2+a^2}\right)=0$, so that the origin of the tortoise coordinate lies on the outer axisymmetric, electrically neutral spherical photon orbit of the EKN geometry.}
\be\label{eq:rast}
	r_{\ast}(r) = r-M -\frac{M^2+a^2}{r-M} +2M\ln\left|\frac{r-M}{\sqrt{M^2+a^2}}\right|
\ee
which is exactly reflected by the geometric inversion
\be
	r-M\ \longrightarrow\ \tilde r-M=\frac{M^2+a^2}{r-M}\,,
	\qquad\Longrightarrow\qquad
	r_\ast(\tilde r)=-r_\ast(r)\,.
\ee
Combining this geometric reflection with the momentum ratio $z$ then yields the CT inversion in the factorized form
\be
	r-M \xrightarrow{\rm CT} \tilde r-M
	=\frac{M^2+a^2}{r-M}\,z
	=\frac{M^2+a^2}{r-M}\,
	\frac{\omega-m\Omega_{\rm H}-q\Phi_{\rm H}}{\omega}\,,
\ee
which is precisely \eqref{eq:CTgeneralExtremal} with $r_{\rm ph}^{(0)}=M$ and $r_1^2=M^2+a^2$.

Finally, supplementing this radial map with the usual conformal transformation of the dependent variable,
\be
	R_{\omega\ell m} \xrightarrow{\text{CT}} \tilde{R}_{\omega\ell m} = \frac{M}{r-M}R_{\omega\ell m} \,,
\ee
one can check directly that the radial equation is mapped into itself, and hence that the CT inversions \eqref{eq:rast} are exact conformal symmetries of the equation of motion.
This confirms that the phase-space dependence of the CT inversion is controlled exactly by the superradiance coefficient, while the remaining part is fixed purely by the requirement of reflecting the tortoise coordinate.

\section{CT symmetry for static perturbations of subextremal Kerr}
\label{sec:SubextremalKerr}

In this section, we exhibit an explicit example of a CT-type inversion symmetry for an asymptotically flat \textit{subextremal} black hole.
We use the results of Ref.~\cite{Lupsasca:2025pnt}, which uncovered a full conformal symmetry governing the static and axisymmetric (spin-$s$) perturbations of the Kerr geometry, i.e., those with $(\omega,m)=(0,0)$.
In this sector, the relevant Teukolsky (and in particular scalar) equations simplify dramatically: after an appropriate change of variables, they become the ordinary flat-space (and in particular Laplace) equation.
This ``trivialization'' makes an inversion symmetry manifest and provides a concrete prototype for how generalized CT inversions can exist beyond extremality, even if their expression in Boyer--Lindquist coordinates is highly nontrivial.

For definiteness, consider a static and axisymmetric scalar field on the Kerr background.
With $x=\cos\theta$ and $\Delta_{r}=r^2-2Mr+a^2$, the wave equation reduces to
\be
    \left[\partial_{r}\Delta_{r}\partial_{r} +\partial_{x}\left(1-x^2\right)\partial_{x}\right]\Phi_{\omega=0,m=0} = 0 \,.
\ee
As recently shown~\cite{Lupsasca:2025pnt}, this equation can be mapped exactly to its flat-space limit as $(M,a)\to(0,0)$,
\be
    \left[\partial_{R}R^2\partial_{R} +\partial_{X}\left(1-X^2\right)\partial_{X}\right]\Phi_{\omega=0,m=0} = 0 \,,
\ee
by introducing the ``trivializing'' coordinates $(R,X)$ in place of Boyer--Lindquist $(r,x)$:
\be\ba
	R\left(r,x\right) &= \mathrm{sign}\left\{r-M\right\}\sqrt{\left(r-M\right)^2-\varkappa^2\left(1-x^2\right)} \,, \\
	X\left(r,x\right) &= \frac{r-M}{R\left(r,x\right)}x \,.
\ea\ee
In the above expressions, we have introduced a quantity
\be
    \varkappa=\sqrt{M^2-a^2} \,,
\ee
which is proportional to the surface gravity of the event horizon of the Kerr black hole, and such that the horizon radii are $r_{\pm}=M\pm\varkappa$.\footnote{Replacing $\varkappa=\sqrt{M^2-a^2-Q^2}$ extends the results of Ref.~\cite{Lupsasca:2025pnt} to subextremal Kerr--Newman.}

In these variables, the inversion symmetry is completely elementary: since the equation is just the flat-space (axisymmetric) Laplacian, it is invariant under the spatial inversion
\be\label{eq:CTNonExtremal}
	R \longrightarrow \tilde{R}=\frac{\alpha^2}{R} \,,\quad
    X\longrightarrow \tilde{X}=X \,,
\ee
for an arbitrary length scale $\alpha$.
Despite its simple appearance in $(R,X)$, this transformation becomes highly nontrivial in Boyer--Lindquist coordinates because $(R,X)$ mix $r$ and $x$ in a nonlinear way.

To make this explicit, we use the inverse map $(R,X)\mapsto(r,x)$:
\be\ba
	r\left(R,X\right)-M &= \mathrm{sign}\left\{R\right\}\sqrt{\frac{R^2+\varkappa^2}{2}\left[1+\sqrt{1-\frac{4\varkappa^2R^2X^2}{\left(R^2+\varkappa^2\right)^2}}\right]} \,, \\
	x\left(R,X\right) &= \frac{R}{r\left(R,X\right)-M}X \,.
\ea\ee
Applying the transformation \eqref{eq:CTNonExtremal} and rewriting the result back in terms of $(r,x)$ yields the CT-like inversion in Boyer--Lindquist form,
\be\ba\label{eq:CTNonExtremalBL}
	r-M &\longrightarrow \tilde{r}-M = \frac{\alpha^2}{R}\sqrt{\frac{1+\nu}{2}\left[1+\sqrt{1-\frac{4\nu X^2}{\left(1+\nu\right)^2}}\right]} \,, \\
	x &\longrightarrow \tilde{x} = \mathrm{sign}\left\{x\right\}\sqrt{\frac{1+\nu}{2\nu}\left[1-\sqrt{1-\frac{4\nu X^2}{\left(1+\nu\right)^2}}\right]} \,,
\ea\ee
where we introduced the dimensionless combination
\be
	\nu = \frac{\varkappa^2R^2}{\alpha^4} = \frac{\varkappa^2}{\alpha^4}\left[\left(r-M\right)^2-\varkappa^2\left(1-x^2\right)\right] \,.
\ee
Thus, even though the underlying symmetry is a simple inversion in the trivializing coordinate $R$, in Boyer--Lindquist variables it becomes a complicated transformation that mixes the radial and angular coordinates.

Two further remarks are worth emphasizing.
First, in the extremal limit $\varkappa\to0$, one has $R\to r-M$, $X\to x$, and $\nu\to0$, and the inversion reduces smoothly to the familiar extremal CT map (up to the arbitrary scale $\alpha$),
\be
    \tilde{r}-M = \frac{\alpha^2}{r-M} +\mathcal{O}\left(\varkappa^2\right) \,,\quad \tilde{x} = x +\mathcal{O}\left(\varkappa^2\right) \,.
\ee
Second, the transformation \eqref{eq:CTNonExtremalBL} shares an important qualitative feature with the conventional CT inversion: it exchanges the event horizon with infinity,
\be
	r\to\infty\quad \longleftrightarrow\quad r=r_+=M+\varkappa\,,
\ee
while mapping the inner horizon to the opposite asymptotic end ($r\to-\infty$) in this coordinate description.
This strongly suggests that the mapping \eqref{eq:CTNonExtremalBL} captures the static $(\omega,m)=(0,0)$ sector of a more general CT-type symmetry, should such a symmetry exist at finite frequency.

Finally, we stress that the photon-shell considerations of Section~\ref{sec:PhotonSpheres} do not directly apply here, since we are working in a strictly static $(\omega\to0)$ and axisymmetric sector rather than the eikonal regime $(\omega\to\infty)$ where wave propagation is well approximated by null geodesic congruences.\footnote{In the high-frequency limit, a massless field is well-approximated by a null geodesic congruence; in particular, eikonal quasinormal modes are congruences of asymptotically bound orbits~\cite{Press:1971wr,Goebel:1972,Ferrari:1984zz,Mashhoon:1985cya,Iyer:1986np,Iyer:1986nq,Seidel:1989bp,Berti:2005eb,Cardoso:2008bp,Hod:2009td,Dolan:2010wr,Yang:2012he,Li:2021zct,Hadar:2022xag,Kapec:2022dvc,Detournay:2025xqd}.}
Moreover, because the trivialized equation is exactly scale-free, the parameter $\alpha$ in \eqref{eq:CTNonExtremal} is not fixed by any intrinsic geometric or scattering data, and there is correspondingly no meaningful notion of superradiant coefficients in this static setting.

Nevertheless, the maps \eqref{eq:CTNonExtremal}--\eqref{eq:CTNonExtremalBL} provide an explicit and highly nontrivial example of a CT-like inversion symmetry acting on an asymptotically flat \emph{subextremal} black hole background.
The main lesson is that generalized CT inversions, if they exist beyond the static sector, may be simple only after passing to appropriately chosen variables, and can appear unwieldy (or even opaque) in standard Boyer--Lindquist coordinates.
This observation motivates the search for analogous ``trivializing'' coordinates in non-static sectors, where one might hope to reconstruct a full generalized CT symmetry at finite $(\omega,m)$.

\section{Conclusion}
\label{sec:Conclusion}

A closer look at the standard arguments against generalized Couch--Torrence (CT) inversions shows that they do not, in fact, provide an obstruction.
The lukewarm Reissner--Nordstr\"om--(A)dS example demonstrates that CT symmetry need neither exchange an event horizon with infinity nor preserve the singularity type of the separated radial equation.
Recasting CT inversions in terms of null geodesics then reveals a robust geometric signature: in all known cases, the inversion preserves the photon region (a photon sphere in the spherically symmetric setting, or a photon shell in the rotating setting), and its fixed point coincides with the corresponding spherical photon orbit(s). 
For asymptotically flat extremal Kerr--Newman, we furthermore showed that the phase-space dependence of the CT map is tightly constrained by scattering data: the inversion factorizes into a geometric part that reflects the tortoise coordinate and a dynamical part controlled by the superradiance coefficient.

Finally, using the ``trivializing'' coordinates of Ref.~\cite{Lupsasca:2025pnt}, we exhibited an explicit CT-type inversion symmetry for static, axisymmetric perturbations of subextremal Kerr (and, similarly, Kerr--Newman).
Although the resulting transformation is remarkably complicated in Boyer--Lindquist coordinates, it becomes the elementary flat-space inversion $R\to\alpha^2/R$ in the appropriate variables and reduces smoothly to the conventional extremal CT map as $\varkappa\to0$.

Taken together, these results suggest that CT symmetry may admit a broader generalization than previously understood.
If a full CT inversion exists at finite frequency for generic subextremal black holes, it will likely be most transparent in suitably adapted variables, and its structure would likely reflect the two lessons emphasized here: it ought to preserve the relevant photon region within each superselection sector, and its phase-space dependence should be linked to (and therefore constrained by) superradiant scattering data.

We close by noting some potential physical implications of generalized CT inversions.
Even in the conventional extremal setting, CT symmetry already relates horizon and asymptotic data in nontrivial ways, for instance through the matching of Aretakis and Newman--Penrose conserved quantities \cite{Bizon:2012we,Lucietti:2012xr,Ori:2013iua,Sela:2015vua,Godazgar:2017igz,Bhattacharjee:2018pqb,Fernandes:2020jto,Agrawal:2025fsv}.
It also underlies refined statements about asymptotic structure, including peeling behavior \cite{Borthwick:2023ovc} and the construction of regular conformal representations of timelike infinity \cite{Lubbe:2013yia}, and it is intimately connected to the vanishing of (electric-type) Love numbers in extremal black holes \cite{Kehagias:2024yzn}.
A broader CT framework---valid beyond extremality or beyond the simplest superselection sectors---could therefore have comparably wide-ranging consequences for black hole perturbation theory, scattering, and holography.

\paragraph{Acknowledgments}
P.C. and L.D. are supported by the European Research Council (ERC) Project 101076737 -- CeleBH.
Views and opinions expressed are however those of the authors only and do not necessarily reflect those of the European Union or the European Research Council.
Neither the European Union nor the granting authority can be held responsible for them.
P.C. and L.D. are also partially supported by the INFN Iniziativa Specifica ST\&FI.
A.L. was supported by NSF grant AST-2307888, the NSF CAREER award PHY-2340457, and the Simons Foundation award SFI-MPS-BH-00012593-09.

\clearpage
\appendix

\section{Geodesics in Kerr--Newman-(A)dS}
\label{app:GeodesicsKNLambda}

In this appendix, we study the geodesic motion associated with a particle carrying an electric charge $q$ that propagates in the background of a Kerr--Newman-(A)dS black hole, whose metric and gauge field in Boyer-Lindquist coordinates $\left(t,r,\theta,\phi\right)$ are
\be\ba\label{eq:KNLambdaMetric1}
	ds^2 &= -\frac{\Delta_{r}}{\Xi^2\Sigma}\left(\ed t-a\sin^2\theta\ed\phi\right)^2 +\frac{\Delta_{\theta}\sin^2\theta}{\Xi^2\Sigma}\left(a\ed t-\left(r^2+a^2\right)\ed\phi\right)^2 +\Sigma\left(\frac{\ed r^2}{\Delta_{r}}+\frac{\ed\theta^2}{\Delta_{\theta}}\right) \,, \\
	A &= -\frac{Qr}{\Xi\Sigma}\left(\ed t-a\sin^2\theta\ed\phi\right) \,,
\ea\ee
with
\be\ba\label{eq:KNLambdaMetric2}
	\Sigma\left(r,\theta\right) &= r^2+a^2\cos^2\theta \,,\quad \Xi = 1 +k\frac{a^2}{\ell^2} \,.
\ea\ee
On-shell, the angular and radial discriminant functions $\Delta_{\theta}$ and $\Delta_{r}$ read
\be\ba\label{eq:KNLambdaMetric3}
	\Delta_{\theta}\left(\theta\right) &= 1+k\frac{a^2}{\ell^2}\cos^2\theta \quad\text{and} \\
	\Delta_{r}(r) &= \left(r^2+a^2\right)\left(1-k\frac{r^2}{\ell^2}\right) -2Mr +Q^2 \,,
\ea\ee
and the parameters $M$, $a$, $Q$, and $\ell$ respectively denote the ADM mass (in $G_{\rm N}=c=1$ units), angular momentum length scale, electric charge\footnote{Here, we are not using the canonical gauge field $A^{\text{can}}$, but instead the rescaled field $A=\frac{\kappa_{\rm N}}{\sqrt{2}}A^{\text{can}}$, with $\kappa_{\rm N}=\sqrt{8\pi G_{\rm N}}$ the gravitational coupling constant, such that the Einstein-Maxwell theory is described by the Lagrangian density
\begin{equation*}
	\mathscr{L} = \frac{\sqrt{-g}}{2\kappa_{\rm N}^2}\left(R-2\Lambda-F_{\mu\nu}F^{\mu\nu}\right) \,.
\end{equation*}
This ensures that the charge length scale $Q$ entering the metric coincides with the electric charge $Q_{\rm e}$ associated with the rescaled field, rather than with that of the canonical field, $Q_{\rm e}^{\text{can}} = \frac{\sqrt{2}}{\kappa_{\rm N}}Q$.} and curvature radius of the AdS ($k=-1$)/dS ($k=+1$) spacetime, with the latter related to the cosmological constant according to $\Lambda = k\frac{3}{\ell^2}$. 
Furthermore, the electrostatic potential $\Phi_{\rm e}$ is the projection of the gauge field along the vector $t^{\mu}\partial_{\mu}=\partial_{t}+\frac{a}{r^2+a^2}\partial_{\phi}$,
\be\label{eq:KNLambda_ElectrostaticPotential}
	\Phi_{\rm e} = -t^{\mu}A_{\mu} = \frac{Qr}{\Xi\left(r^2+a^2\right)} \,,
\ee
with the nice property of being a purely radial function.

The effective Lagrangian associated with electrically charged geodesics is
\be
	\mathcal{L} = \frac{1}{2}g_{\mu\nu}\left(x\right)\dot{x}^{\mu}\dot{x}^{\nu} +qA_{\mu}\left(x\right)\dot{x}^{\mu} \,,
\ee
where $\dot{x}^{\mu} = \frac{dx^{\mu}}{ds}$, with $s$ an affine parameter. The conjugate momentum is then
\be
	p_{\mu} = \frac{\partial\mathcal{L}}{\partial\dot{x}^{\mu}} = g_{\mu\nu}\dot{x}^{\nu} +qA_{\mu} \,.
\ee
This results in the Hamiltonian
\be
	\mathcal{H} = p_{\mu}\dot{x}^{\mu} -\mathcal{L} = \frac{1}{2}\left(p_{\mu}-qA_{\mu}\right)\left(p^{\mu}-qA^{\mu}\right) \,.
\ee

\paragraph{Constants of motion}\hfill\newline
Since the background is stationary and axisymmetric, with the associated Killing vectors being $\partial_{t}$ and $\partial_{\phi}$, the energy and angular momentum are constants of motion,
\begin{subequations}\label{eq:KNLambdaELz}
	\begin{align}
		\begin{split}
			{}&E = -p_{t} = \frac{1}{\Xi^2\Sigma}\left\{ \left(\Delta_{r}-a^2\Delta_{\theta}\sin^2\theta\right)\dot{t} -a\sin^2\theta\left(\Delta_{r}-\left(r^2+a^2\right)\Delta_{\theta}\right)\dot{\phi} +\Xi qQr\right\} \,,
		\end{split} \\
		\begin{split}
			{}&L_{z} = p_{\phi} = \frac{\sin^2\theta}{\Xi^2\Sigma}\left\{ a\left(\Delta_{r}-\left(r^2+a^2\right)\Delta_{\theta}\right)\dot{t} +\left(\left(r^2+a^2\right)^2\Delta_{\theta}-a^2\Delta_{r}\sin^2\theta\right)\dot{\phi} +a\Xi qQr \right\} \,.
		\end{split}
	\end{align}
\end{subequations}

On top of that, the Hamiltonian is identified with the (unit) inertial mass-squared
\be
	\mathcal{H} = -\frac{\sigma}{2} \quad\Longrightarrow\quad \sigma = -\left(p_{\mu}-qA_{\mu}\right)\left(p^{\mu}-qA^{\mu}\right) = -g_{\mu\nu}\dot{x}^{\mu}\dot{x}^{\nu}
\ee
where
\be\ba
	\sigma = \begin{cases}
		+1 & \text{for timelike geodesics (massive particles)} \,; \\
		0 & \text{for null geodesics (massless particles)} \,; \\
		-1 & \text{for spacelike geodesics (tachyonic particles)} \,.
	\end{cases}
\ea\ee

Lastly, the KN-(A)dS spacetime enjoys a hidden symmetry generated by a rank-$2$ Killing tensor~\cite{Penrose:1973um,Floyd:1974}, i.e., a symmetric rank-$2$ tensor $K^{\mu\nu}=K^{(\mu\nu)}$ that satisfies $\nabla_{(\rho}K_{\mu\nu)}=0$.
Specifically, its Killing tensor reads
\be\ba
	K^{\mu\nu}\partial_{\mu}\partial_{\nu} &= r^2\partial^2 -\Delta_{r}\,\partial_{r}^2 +\frac{\Xi^2}{\Delta_{r}}\left(\left(r^2+a^2\right)\partial_{t}+a\,\partial_{\phi}\right)^2 \\
	&= -a^2\cos^2\theta\,\partial^2 +\Delta_{\theta}\,\partial_{\theta}^2 +\frac{\Xi^2}{\Delta_{\theta}}\left(a\sin\theta\,\partial_{t}+\frac{1}{\sin\theta}\,\partial_{\phi}\right)^2 \,,
\ea\ee
and gives rise to an additional constant of motion: the quadratic charge $\mathcal{K}=K^{\mu\nu}p_{\mu}p_{\nu}$~\cite{Carter:1968rr}, whose explicit expression is~\cite{Garnier:2023lph}
\be\ba\label{eq:KNLambdaCarterConst}
	\mathcal{K} = K^{\mu\nu}p_{\mu}p_{\nu} &= \frac{1}{\Delta_{\theta}}\left[\Sigma^2\dot{\theta}^2 +\Xi^2\left(aE\sin\theta-\frac{L_{z}}{\sin\theta}\right)^2\right] +\sigma a^2\cos^2\theta \\
	&= \frac{1}{\Delta_{r}}\left[-\Sigma^2\dot{r}^2 +\Xi^2\left(\left(r^2+a^2\right)E-aL_{z}-\frac{qQr}{\Xi}\right)^2\right] -\sigma r^2 \,.
\ea\ee
The conventional Carter constant $\mathcal{C}$~\cite{Carter:1968rr} is then given by the combination
\be
	\mathcal{C}=\mathcal{K}-\Xi^2\left(aE-L_{z}\right)^2 \,,
\ee
which is defined such that $\mathcal{C}=0$ on the equatorial plane $\theta=\frac{\pi}{2}$.

\paragraph{System of geodesics equations}\hfill\newline
The expressions for the energy and angular momentum \eqref{eq:KNLambdaELz} can be immediately inverted to obtain $\dot{t}$ and $\dot{\phi}$.
Similarly, Eq.~\eqref{eq:KNLambdaCarterConst}, along with the conservation of the inertial mass $\sigma=-g_{\mu\nu}\dot{x}^{\mu}\dot{x}^{\nu}$, can be inverted to find separable expressions for $\dot{\theta}$ and $\dot{r}$.
Summarizing this process, one ends up with the following integrable system of geodesic equations~\cite{Carter:1968rr,Garnier:2023lph}
\be\ba\label{eq:KNLambdaGeodesicsEqs}
	\frac{\Sigma}{\Xi E}\dot{r} &= \pm_{r}\sqrt{\mathcal{R}(r)} \,, \\
	\frac{\Sigma}{\Xi E}\dot{\theta} &= \pm_{\theta}\sqrt{\Theta\left(\theta\right)} \,, \\
	\frac{\Sigma}{\Xi^2E}\dot{t} &= \frac{r^2+a^2}{\Delta_{r}}\left(r^2+a^2-ab-\mathfrak{q}r\right) +\frac{a}{\Delta_{\theta}}\left(b-a\sin^2\theta\right) \,, \\
	\frac{\Sigma}{\Xi^2E}\dot{\phi} &= \frac{a}{\Delta_{r}}\left(r^2+a^2-ab-\mathfrak{q}r\right) +\frac{1}{\Delta_{\theta}\sin^2\theta}\left(b-a\sin^2\theta\right) \,,
\ea\ee
where the effective potentials $\mathcal{R}(r)$ and $\Theta\left(\theta\right)$ are given by
\be\ba
	\mathcal{R}(r) &= \left(r^2+a^2-ab-\mathfrak{q}r\right)^2 -\Delta_{r}\left(\eta+\left(a-b\right)^2+\mu r^2\right) \,, \\
	\Theta\left(\theta\right) &= -\left(a\sin\theta-\frac{b}{\sin\theta}\right)^2  +\Delta_{\theta}\left(\eta+\left(a-b\right)^2-\mu a^2\cos^2\theta\right) \,,
\ea\ee
and we have introduced the impact parameter $b$, the geodesic charge length scale $\mathfrak{q}$, the rescaled Carter constant $\eta$ (with units of length-squared) and the dimensionless mass-squared parameter $\mu$ according to
\be
	b = \frac{L_{z}}{E} \,,\quad \mathfrak{q} = \frac{qQ}{\Xi E} \,,\quad \eta = \frac{\mathcal{C}}{\Xi^2E^2} \quad\text{and}\quad \mu = \frac{\sigma}{\Xi^2E^2} \,,
\ee
while $\pm_{r}= \mathrm{sign}\left\{\dot{r}\right\}$ and $\pm_{\theta}= \mathrm{sign}\left\{\dot{\theta}\right\}$ are a shorthand notation for the signs of the corresponding $4$-velocity.

\subsection{Photon shells}

Spherical orbits are located at constant radii $r=r_{\circ}=\text{const.}$ for which both the radial velocity and radial acceleration are zero, $\dot{r}=0$ and $\ddot{r}=0$.
From the system~\eqref{eq:KNLambdaGeodesicsEqs} of geodesic equations, one then sees that this reduces to imposing the conditions\footnote{More explicitly, the radial acceleration can be worked out from Eq.~\eqref{eq:KNLambdaGeodesicsEqs} to be
\begin{equation*}\ba
    \ddot{r} &= \partial_{r}\left(\frac{\Xi^2E^2\mathcal{R}}{\Sigma^2}\right) + \partial_{\theta}\left(\frac{\Xi^2E^2\mathcal{R}}{\Sigma^2}\right)\frac{\dot{\theta}}{\dot{r}}
    = \partial_{r}\left(\frac{\Xi^2E^2\mathcal{R}}{\Sigma^2}\right) \pm_{r}\pm_{\theta}\frac{4a^2\sin\theta\cos\theta}{\Sigma^3}\Xi^2E^2\sqrt{\mathcal{R}\Theta} \,.
\ea\end{equation*}
For spherical orbits, $\dot{r}=0$ sets the second term automatically to zero and the $\ddot{r}=0$ conditions outputs the additional condition that $\partial_{r}\left(\frac{\Xi^2E^2\mathcal{R}}{\Sigma^2}\right)\bigg|_{r=r_{\circ}}$.}
\be\label{eq:SphericalOrbitsConditionsFull}
    \left(\frac{\Xi^2E^2\mathcal{R}}{\Sigma^2}\right)\bigg|_{r=r_{\circ}} = 0 \quad\text{and}\quad \partial_{r}\left(\frac{\Xi^2E^2\mathcal{R}}{\Sigma^2}\right)\bigg|_{r=r_{\circ}} = 0 \,.
\ee
Spherical orbits for which $\Sigma$ does not blow up (i.e., away from infinity) and whose orbital energy $E$ is non-zero are double roots of the effective radial potential,
\be\label{eq:SphericalOrbitsConditions}
	\mathcal{R}\left(r_{\circ}\right) = 0 \,,\quad \mathcal{R}^{\prime}\left(r_{\circ}\right) = 0 \,.
\ee
This gives rise to a collection of spherical orbits that depend solely on the phase space variables $\left(b,\eta\right)$, once the mass and electric charge of the particle, and also the parameters $\left(\frac{k}{\ell^2},a,Q\right)$ that characterize the background geometry, are specified,
\be
	r_{\circ} = r_{\circ}\left(b,\eta\right) \,.
\ee
While the radius of such orbits is constant, the polar angle $\theta$ can in general oscillate between turning points $\theta_{\pm}$,
\be
	\Theta\left(\theta\right)\ge0 \quad\Longrightarrow\quad \theta_{-}\left(b,\eta\right) \le \theta\le \theta_{+}\left(b,\eta\right) \,\quad\text{with}\quad \Theta\left(\theta_{\pm}\right)=0 \,,
\ee
due to the breaking of spherical symmetry from the angular momentum of the background configuration.
Nevertheless, axisymmetry of the geometry still guarantees degeneracy of motion along the azimuthal circle and, furthermore, ensures reflection symmetry with respect to the equatorial plane $\theta=\frac{\pi}{2}$.

Spherical photon orbits $r_{\rm ph}$, in particular, are the spherical orbits associated with null geodesics ($\mu=0$), $r_{\rm ph} = r_{\circ}^{\left(\mu=0\right)}$.
The conditions $\mathcal{R}\left(r_{\rm ph}\right) = 0 = \mathcal{R}^{\prime}\left(r_{\rm ph}\right)$ can be solved explicitly to write down the relation between a particular spherical photon orbit location and the values of the phase space variables $\left(b,\eta\right)$~\cite{Wilkins:1972rs,Gralla:2019ceu,Gralla:2019xty,Teo:2003ltt,Teo:2020sey}, namely,
\be\ba\label{eq:KNLambdaImpactCarterNull}
	ab &= \left[r^2+a^2-\mathfrak{q}r-\frac{2\Delta_{r}}{\Delta_{r}^{\prime}}\left(2r-\mathfrak{q}\right)\right]\bigg|_{r=r_{\rm ph}} \,, \\
	\eta &= \left[\frac{4\Delta_{r}}{\Delta_{r}^{\prime2}}\left(2r-\mathfrak{q}\right)^2\right]\bigg|_{r=r_{\rm ph}} -\left(a-b\right)^2 \,.
\ea\ee

Now, the range of the phase space variables $\left(b,\eta\right)$ must in general be such that the geodesic motion is real, i.e., such that both the radial and the angular effective potentials are positive semi-definite,
\be
	\mathcal{R}(r) \ge0 \quad\text{and}\quad \Theta\left(\theta\right)\ge0 \,.
\ee

\paragraph{Phase space constraints from angular motion} \hfill \\
To analyze the angular motion of null geodesics, first change coordinates to
\be
	u= \cos^2\theta \,,
\ee
and introduce quantities
\be\ba
	\alpha_{u} &= -a^2\left(1+k\frac{\eta+\left(a-b\right)^2}{\ell^2}\right) \quad\text{and} \\
    \beta_{u} &= 2a\left(a-b\right) -\left(1-k\frac{a^2}{\ell^2}\right)\left(\eta+\left(a-b\right)^2\right) \,.
\ea\ee
Then, the condition $\Theta\left(\theta\right)\ge0$ is equivalent to
\be
	\alpha_{u}u^2 +\beta_{u}u +\eta \ge0 \,.
\ee

\begin{figure}
	\centering
	\includegraphics[width=0.5\textwidth]{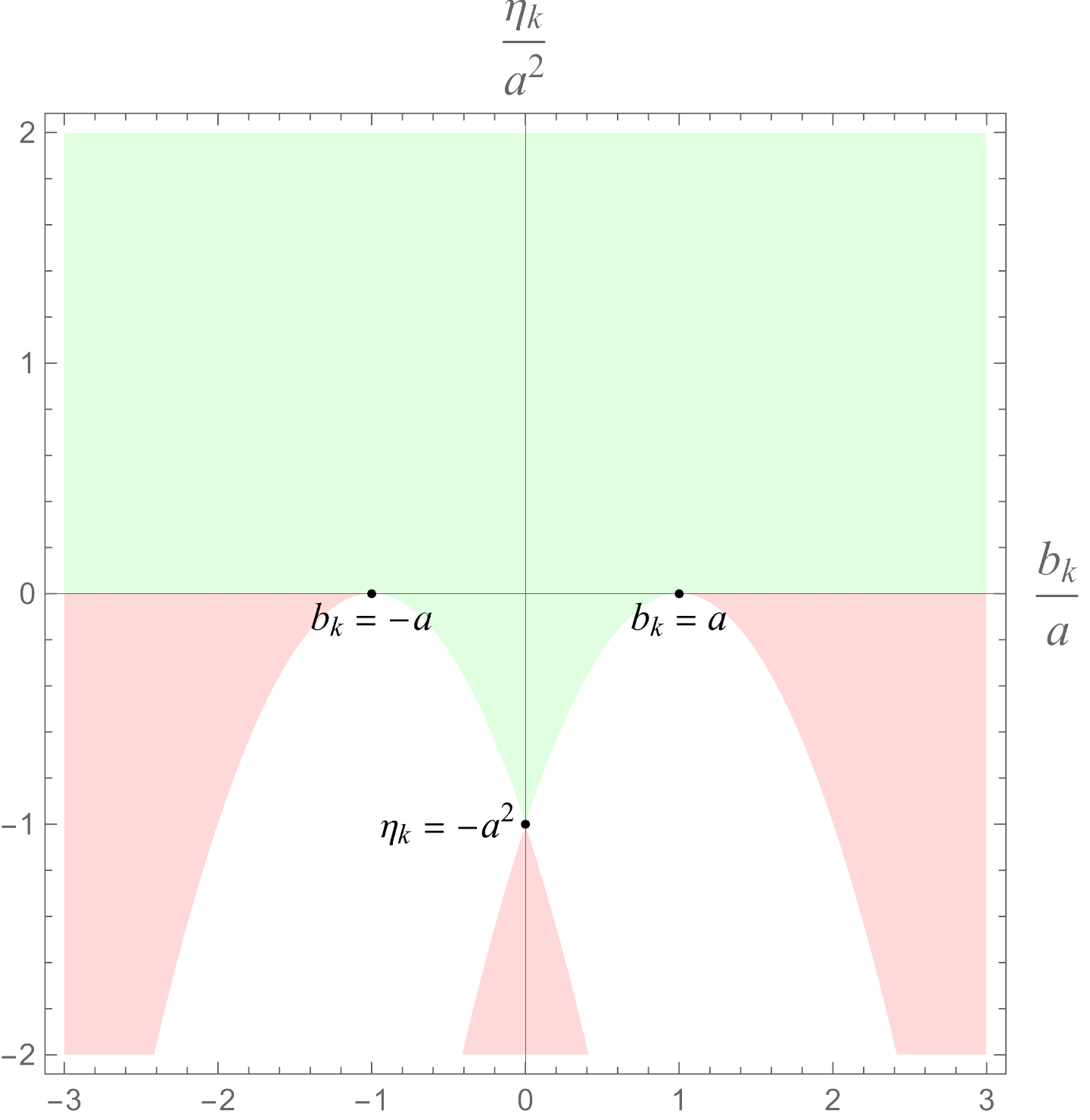}
	\caption{Allowed range (green region) of redefined Carter constant $\eta_{k}$ and impact parameter $b_{k}$, related to the original (rescaled) Carter constant $\eta$ and impact parameter $b=\frac{L_{z}}{E}$ bia Eq.~\eqref{eq:EtaB_EtakBk}.
    The red region spans values of $\left(b_{k},\eta_{k}\right)$ for which the turning points $u_{\pm}$ of the angular motion are still real-valued, but violate the $u_{-}\le1$ or $u_{+}\ge0$ conditions.
    The white region corresponds to complex-valued $u_{\pm}$.}
	\label{fig:etabAllowedKNAdS}
\end{figure}

When $\alpha_{u}\le0$,\footnote{$\alpha_{u}\le0$ is equivalent to $k\frac{\eta+\left(a-b\right)^2}{\ell^2}\ge-1$, which always holds for asymptotically flat ($k=0$) or dS ($k=+1$) spacetimes, but requires $\eta+\left(a-b\right)^2\le\ell^2$ in asymptotically AdS ($k=-1$) spacetimes.} polar motion is only allowed between the turning points $u_{\pm}$ at
\be\ba
	u_{\pm} &= u_0 \pm\sqrt{u_0^2 +\frac{\eta}{a^2\left(1+k\frac{\eta+\left(a-b\right)^2}{\ell^2}\right)}} \,,\quad\text{with} \\
	&u_0= -\frac{\beta_{u}}{2\alpha_{u}} = \frac{2a\left(a-b\right)-\left(1-k\frac{a^2}{\ell^2}\right)\left(\eta+\left(a-b\right)^2\right)}{2a^2\left(1+k\frac{\eta+\left(a-b\right)^2}{\ell^2}\right)} \,.
\ea\ee
In particular, polar motion is only possible when
\be\label{eq:upmAngular}
	u_{+}\ge0 \quad\text{and}\quad u_{-}\le 1 \,.
\ee
To analyze these constraints, introduce a new Carter constant and impact parameter
\be\label{eq:EtaB_EtakBk}
	\eta_{k} = \frac{\eta}{1+k\frac{\eta+\left(a-b\right)^2}{\ell^2}} \,,\quad b_{k} = \frac{b}{\sqrt{1+k\frac{\eta+\left(a-b\right)^2}{\ell^2}}} \,,
\ee
in terms of which the turning points of the angular potential take the form
\be
	u_{\pm} = u_0 \pm \sqrt{u_0^2 +\frac{\eta_{k}}{a^2}} \,,\quad
    u_0 = \frac{1}{2}\left(1-\frac{\eta_{k}+b_{k}^2}{a^2}\right) \,.
\ee
These formulas are identical to their counterparts in the asymptotically flat Kerr--Newman geometry, provided that we identify $\eta_{k=0}=\eta$ and $b_{k=0}=b$~\cite{Gralla:2019ceu}.
Using this identification, the values of $b_{k}$ and $\eta_{k}$ allowed by Eq.~\eqref{eq:upmAngular} are (see Fig.~\ref{fig:etabAllowedKNAdS})~\cite{Gralla:2019ceu}
\be\label{eq:NullGeodesicsAllowed1}
	\eta_{k} \ge\begin{cases}
		0 & \text{if $\left|b_{k}\right|\ge \left|a\right|$} \,; \\
		-\left(\left|b_{k}\right|-\left|a\right|\right)^2 & \text{if $\left|b_{k}\right|\le \left|a\right|$} \,.
	\end{cases}
\ee
The subsequent phase-space constraints on the original variables $b$ and $\eta$ can then be extracted from the inverse relation between $\left(b,\eta\right)$ and $\left(b_{k},\eta_{k}\right)$,
\be
	\eta = \eta_{k}\frac{1+k\frac{\left(a-b\right)^2}{\ell^2}}{1-k\frac{\eta_{k}}{\ell^2}} \quad\text{with}\quad b = \frac{b_{k}}{1-k\frac{\eta_{k}+b_{k}^2}{\ell^2}}\left[\sqrt{\Xi-k\frac{\Xi\eta_{k}+b_{k}^2}{\ell^2}} -k\frac{ab_{k}}{\ell^2}\right] \,.
\ee

\paragraph{Photon shells of extremal Kerr--Newman black holes}\hfill\\
We now extract the spherical photon orbits, whose existence also constrains the range of $\left(b,\eta\right)$, focusing to the particular geometry of the asymptotically flat extremal Kerr--Newman black hole, for which $k=0$ and $a^2+Q^2=M^2$, and hence $\Delta_{r}(r)=\left(r-M\right)^2$.
In general, Eq.~\eqref{eq:KNLambdaImpactCarterNull} gives rise to a fourth-order algebraic equation involving $b$ and one sixth-order algebraic equation involving $\eta$.
For the EKN geometry, however, Eq.~\eqref{eq:KNLambdaImpactCarterNull} greatly simplifies to two second-order algebraic equations,
\be\ba\label{eq:EKNImpactCarterNull}
	{}&r_{\rm ph}^2-2Mr_{\rm ph} +M\mathfrak{q} +a\left(b-a\right) = 0 \,, \\
	&\left(2r_{\rm ph}-\mathfrak{q}\right)^2 - \eta-\left(a-b\right)^2 = 0 \,.
\ea\ee
The first one implies that for each allowed value of the impact parameter, there are two spherical photon orbits,
\be
	r_{\rm ph}^{\left(\pm\right)}(b) = M\pm\sqrt{M^2+a^2-ab-\mathfrak{q}M} \,.
\ee
The outer spherical photon orbit, $r_{\rm ph}^{\left(+\right)}(b)$, is always in the black hole exterior, and hence timelike, while $r_{\rm ph}^{\left(-\right)}(b)$ is always in the interior, and therefore spacelike.
Each collection of spherical photon orbits then makes up a three-dimensional surface of bound null geodesics, known as a photon shell \cite{Johnson:2019ljv}.
The fact that the photon shells have finite thickness comes from the following observations:
\begin{itemize}
	\item The angular geodesic motion implies that the allowed values of the phase space parameters must live in the green region of Fig.~\ref{fig:etabAllowedKNAdS}, spelled out in Eq.~\eqref{eq:NullGeodesicsAllowed1}~\cite{Gralla:2019ceu}.
	\item The circularity of the radial geodesic motion implies Eqs.~\eqref{eq:EKNImpactCarterNull}, which define curves $\eta(b)$.
    A photon shell is spanned by values of the impact parameter for which the corresponding curve lays in the green region of the phase space diagram Fig.~\ref{fig:etabAllowedKNAdS}.
\end{itemize}

\begin{figure}
	\centering
	\begin{subfigure}[t]{0.5\textwidth}
		\centering
		\includegraphics[width=\textwidth]{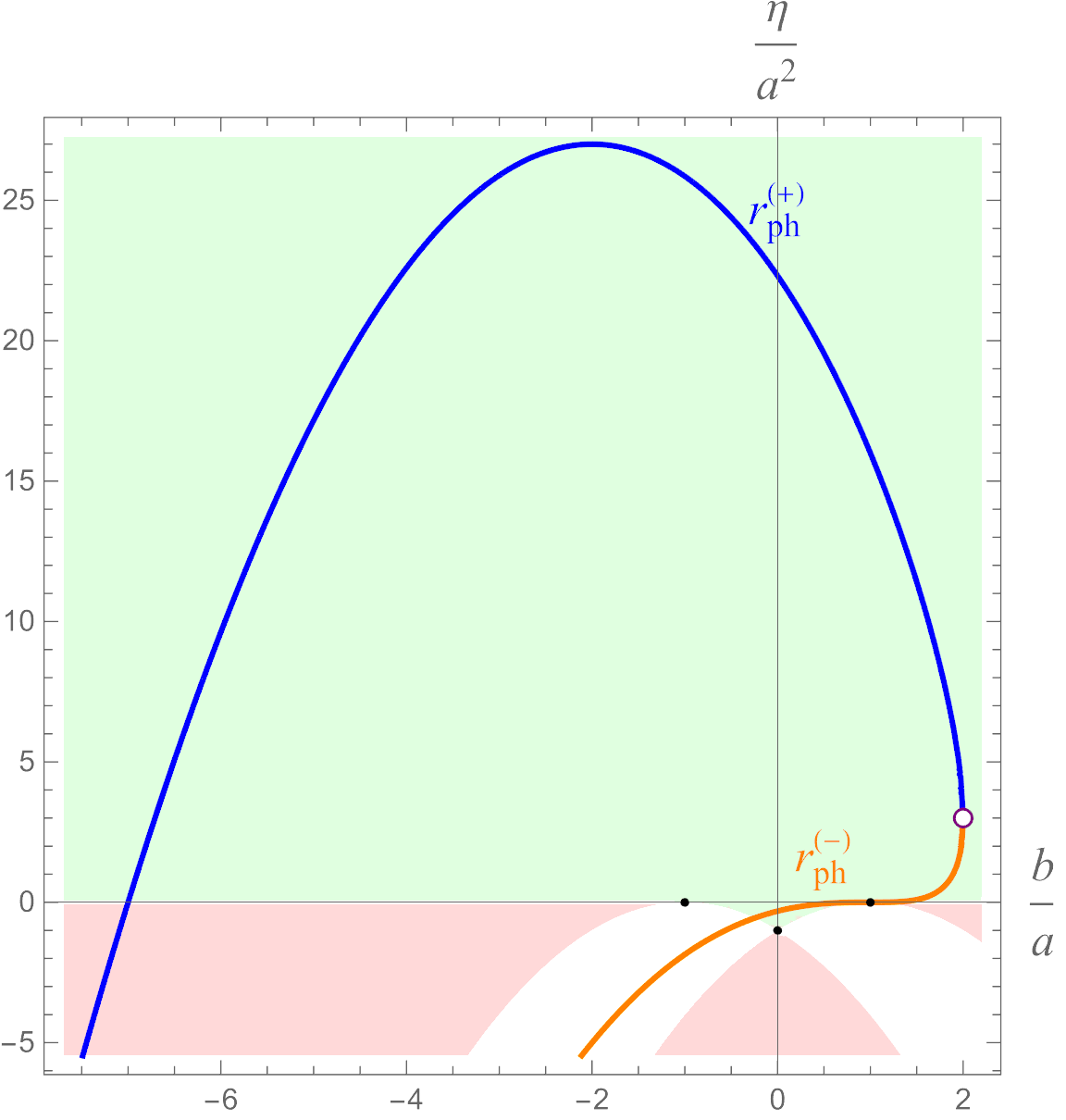}
		\caption{}
		\label{fig:etabAllowedKNAdS_ExtremalKerrq0}
	\end{subfigure}~
	\begin{subfigure}[t]{0.5\textwidth}
		\centering
		\includegraphics[width=\textwidth]{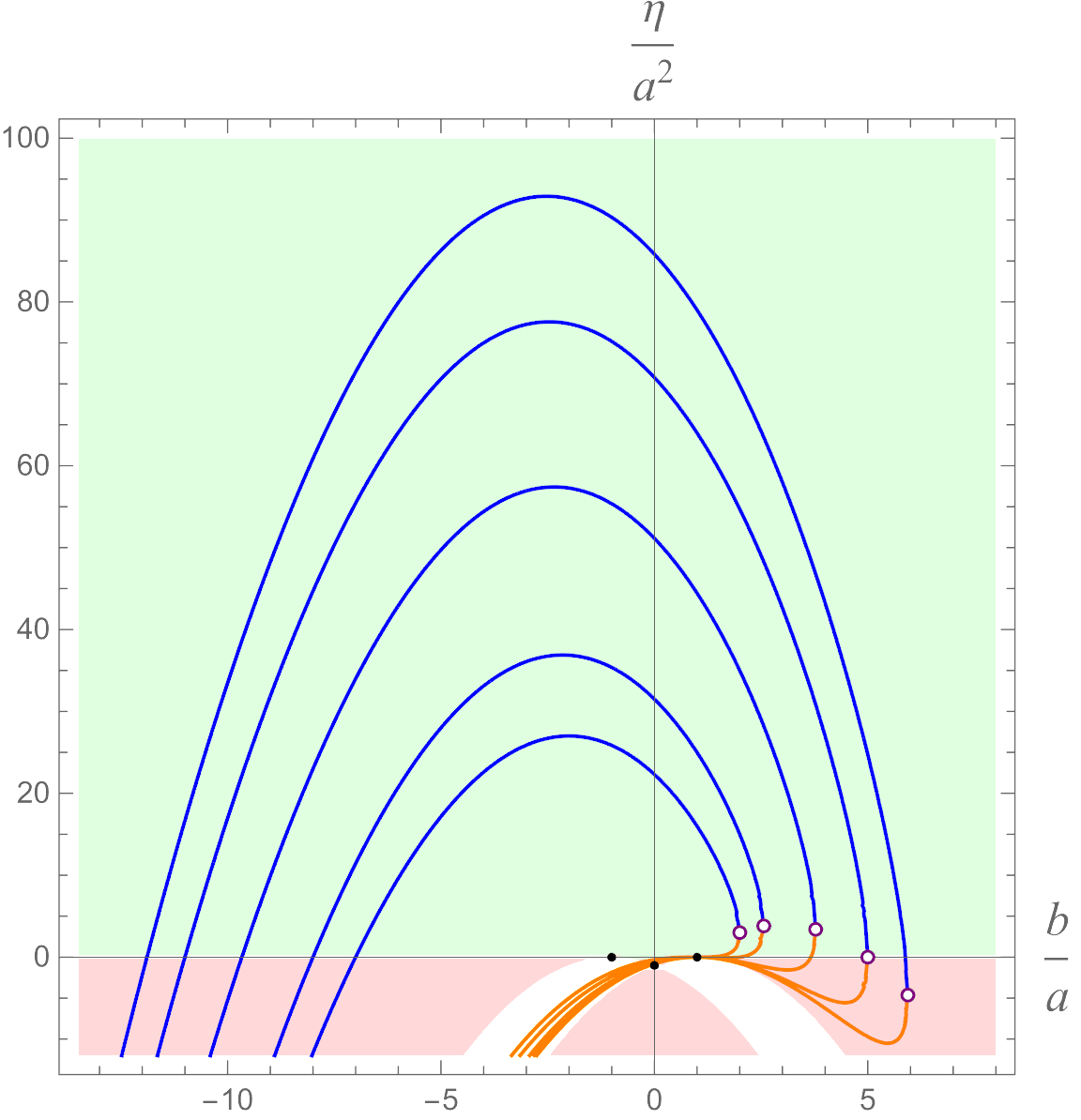}
		\caption{}
		\label{fig:etabAllowedKNAdS_ExtremalKerrNewmanq0Q}
	\end{subfigure}
	\caption{Phase space diagram, in the $\left(b,\eta\right)$ space, demonstrating the electrically neutral ($\mathfrak{q}=0$) inner and outer photon shells of the clock-wisely rotating extremal Kerr--Newman black hole ($a=+\sqrt{M^2-Q^2}$), for various values of the charge parameter $Q$: (a) for $Q=0$ (extremal Kerr); (b) starting from the innermost set of curves, $\left|a\right|/M=1.0, 0.8, 0.6, 0.5, 0.45$, corresponding to $\left|Q\right|/M= 0.0, 0.6, 0.8, 0.866, 0.893$ respectively. The blue line represents the curve associated with outer spherical photon orbits $r_{\rm ph}^{\left(+\right)}$, while the orange line represents the curve associated with inner spherical photon orbits $r_{\rm ph}^{\left(-\right)}$. The corresponding photon shells are the parts of these curves that reside in the green region (phase space allowed from angular motion).}
	\label{fig:etabAllowedKNAdS_ExtremalKerrNewmanq0}
\end{figure}

For instance, the outer photon shell associated with electrically charged null geodesics in the background of an EKN black hole are spanned by
\be
	\text{\underline{For outer photon shell of EKN}: }\quad \left(ab\right)_{\text{min}}\le ab< \left(ab\right)_{\text{max}} \,,
\ee
with
\be\ba
	\left(ab\right)_{\text{max}} &=
	\begin{cases}
		M^2+a^2-\mathfrak{q}M & \text{if $\frac{\left|a\right|}{M}\ge\frac{M-\mathfrak{q}}{2M-\mathfrak{q}}$} \,, \\
		\left|a\right|\left(\sqrt{M-\left|a\right|}+\sqrt{M-\left|a\right|-\mathfrak{q}}\right)^2 & \text{if $\frac{\left|a\right|}{M}<\frac{M-\mathfrak{q}}{2M-\mathfrak{q}}$} \,,
	\end{cases} \\
	\left(ab\right)_{\text{min}} &= -\left|a\right|\left(\sqrt{M+\left|a\right|}+\sqrt{M+\left|a\right|-\mathfrak{q}}\right)^2 \,.
\ea\ee
The upper bound of $ab< M^2+a^2-\mathfrak{q}M$ follows from the requirement that there exist spherical photon orbits, $r_{\rm ph}^{\left(\pm\right)}(b)\in\mathbb{R}$.
The other bounds come from the phase space constraints of the angular motion, Eq.~\eqref{eq:NullGeodesicsAllowed1}, that is, they correspond to the points where the exterior spherical photon orbit curve $r_{\rm ph}^{\left(+\right)}$ intersects the $b$-axis.
For $\frac{\left|a\right|}{M}\ge\frac{M-\mathfrak{q}}{2M-\mathfrak{q}}$, $r_{\rm ph}^{\left(+\right)}$ intersects the $b$-axis only once, imposing the lower bound on $\left(ab\right)_{\text{min}}$ above, while for sufficiently slow rotation (namely, for $\frac{\left|a\right|}{M}<\frac{M-\mathfrak{q}}{2M-\mathfrak{q}}$), the exterior spherical photon orbit curve intersects the $b$-axis twice, with the new intersection point dominating the upper bound $\left(ab\right)_{\text{max}}$ as well.
This picture is illustrated in Fig.~\ref{fig:etabAllowedKNAdS_ExtremalKerrNewmanq0} for an electrically neutral ($\mathfrak{q}=0$) massless particle, for which
\be\ba
	\left(ab\right)_{\text{max}} &=
	\begin{cases}
		M^2+a^2 & \text{if $\frac{\left|a\right|}{M}\ge\frac{1}{2}$} \,, \\
		\left|a\right|\left(4M-3\left|a\right|\right) & \text{if $\frac{\left|a\right|}{M}<\frac{1}{2}$} \,,
	\end{cases} \\
	\left(ab\right)_{\text{min}} &= -\left|a\right|\left(4M+3\left|a\right|\right) \,. \\	
\ea\ee
In particular, for the extremal Kerr black hole, for which $\left|a\right|=M$, this agrees with the known result that $\left(ab\right)_{\text{min}}=-7M^2$ and $\left(ab\right)_{\text{max}}=2M^2$~\cite{Gralla:2019ceu,Gralla:2019drh}; see Fig.~\ref{fig:etabAllowedKNAdS_ExtremalKerrq0}.

As for the inner photon shell, an analogous analysis shows that
\be\ba
	{}&\text{\underline{For inner photon shell of EKN}: } \quad -\left(\frac{a^2-\frac{\mathfrak{q}^2}{4}}{2M-\mathfrak{q}}\right)^2\le ab\le a^2 \quad\text{OR} \\
	&\begin{cases}
		\left|a\right|\left(\sqrt{M-\left|a\right|}+\sqrt{M-\left|a\right|-\mathfrak{q}}\right)^2\le ab< M^2+a^2-\mathfrak{q}M & \text{if $\frac{\left|a\right|}{M}\ge\frac{M-\mathfrak{q}}{2M-\mathfrak{q}}$} \,, \\
		\varnothing & \text{if $\frac{\left|a\right|}{M}<\frac{M-\mathfrak{q}}{2M-\mathfrak{q}}$} \,.
	\end{cases}
\ea\ee

Finally, the extremal Kerr--Newman geometry possesses one more collection of spherical photon orbits, spanned by stable spherical photon orbits $r_{\rm ph}^{\left(0\right)}=M$ located exactly on the degenerate horizon, for which $ab=M^2+a^2-\mathfrak{q}M$.
These orbits exist for all values of the Carter constant, and hence fill up an entire photon sphere.

\addcontentsline{toc}{section}{References}
\bibliographystyle{JHEP}
\bibliography{CTGeneralized}

\end{document}